\documentclass[epj]{svjour}
\usepackage{amsmath}
\usepackage{amssymb}
\usepackage{epsfig}
\newcommand{\repizeropm}{\mathrm{Re}\mspace{2.5mu}^0\!\Pi^{+-}}
\newcommand{\impizeropm}{\mathrm{Im}\mspace{2.5mu}^0\!\Pi^{+-}}
\newcommand{\impipm}{\mathrm{Im}\mspace{2.5mu}\Pi^{+-}}
\newcommand{\repipm}{\mathrm{Re}\mspace{2.5mu}\Pi^{+-}}
\newcommand{\pisigsig}{\Pi^{\sigma -\!\sigma}}
\newcommand{\pisigsigzero}{^0\!\Pi^{\sigma -\!\sigma}}
\newcommand{\tast}{t_\ast}
\newcommand{\x}{x=\frac{1}{2}U|\mu|}
\newcommand{\quasi}{Z=[1-(\partial\Sigma_\uparrow^{\mathrm{R}}(\omega)/\partial \omega)_{\omega=0}]^{-1}}
\newcommand{\mj}{M. Jarrell}
\newcommand{\tp}{T. Pruschke}
\newcommand{\del}{D. E. Logan}
\newcommand{\rb}{R. Bulla}
\newcommand{\prb}{Phys. Rev. B }
\newcommand{\prl}{Phys. Rev. Lett. }
\newcommand{\jpcm}{J. Phys. Condens. Matt. }
\newcommand{\nld}{N. L. Dickens}
\newcommand{\mtg}{M. T. Glossop}
\newcommand{\rmp}{Rev. Mod. Phys. }
\newcommand{\nsv}{N. S. Vidhyadhiraja}
\newcommand{\hrk}{H. R. Krishnamurthy}
\numberwithin{equation}{section}
\begin{document}
\title{Kondo insulators in the periodic Anderson model: a local moment approach}
\titlerunning{Kondo Insulators: a local moment approach}
\authorrunning{Victoria E. Smith {\it{et al.}}}
\author{Victoria\ E.\ Smith\inst{1}, David E.\ Logan\inst{1} \and H.\ R.\ Krishnamurthy\inst{2}
}                     
%
%
\institute{Oxford University, Physical and Theoretical Chemistry, South Parks Road, Oxford OX1 3QZ, UK. \and Department of Physics, I. I. Sc., Bangalore  560 012; and JNCASR, Jakkur, Bangalore 506 064, India.}
\date{Received: date / Revised version: date}
%
\abstract{The symmetric periodic Anderson model is well known to capture the essential physics of Kondo insulator materials. Within the framework of dynamical mean-field theory, we develop a local moment approach to its single-particle dynamics in the paramagnetic phase. The approach is intrinsically non-perturbative, encompasses all energy scales and interaction strengths, and satisfies the low-energy dictates of Fermi liquid theory. It captures in particular the strong coupling behaviour and exponentially small quasiparticle scales characteristic of the Kondo lattice regime, as well as simple perturbative behaviour in weak coupling. Particular emphasis is naturally given to strong coupling dynamics, where the resultant clean separation of energy scales enables the scaling behaviour of single-particle spectra to be obtained.
} 
\maketitle
\section{Introduction}
\label{sec:intro}
In the field of heavy fermion physics, the periodic Anderson model (PAM) has long played a key role in understanding the rich range of behaviour arising in these lanthanide or actinide based compounds \cite{1,2}. A natural lattice generalization of the single-impurity Anderson model \cite{2}, {\it each} lattice site in the PAM contains a non-degenerate, correlated f-orbital that hybridises locally to a non-\hspace{-0.2mm}interacting conduction band. The model is certainly a simplification of material reality, albeit rather a good one for $Ce$-based systems where crystal field splittings reduce the multiplicity of the $\mathrm{f}^1$-configuration to a Kramers doublet. Yet its simplicity is nominal: a microscopic understanding of the PAM is far from complete, and even in low-temperature Fermi liquid phases many basic issues relating to the formation, nature and description of coherent quasiparticles remain open. 

In recent years, considerable progress in understanding correlated electron systems has been made within the powerful framework of dynamical mean-field theory (DMFT, reviewed in \cite{3,4,5,6}); which is formally exact in the limit of infinite spatial dimensions, and provides a tangible local approximation in finite dimensions without trivialising the central role of interactions. Within DMFT the dynamics of the system become essentially local, and all correlated lattice-fermion models reduce to an effective quantum impurity hybridizing self-consistently with the surrounding fermionic bath \cite{3,4,5,6}. But therein lies an unsurprising difficulty, for to solve such problems entails the ability to describe an Anderson impurity model (AIM) with essentially arbitrary dynamics ($\omega$-dependence) in the hybridization function $\Delta_{\mathrm{eff}}(\omega)$ that encodes the coupling between the fiducial impurity and the underlying host/bath. Such a description should be capable of (i) handling the problem on {\it all} energy scales and (ii) across the {\it full} range of (f-electron) interaction strengths U; and (iii) at low-energies must satisfy the dictates of Fermi liquid theory arising from adiabatic continuity to the non-interacting limit (assuming such a state arises). As such it must be intrinsically non-perturbative in order to capture the exponentially small scales symptomatic of strongly correlated behaviour, yet must also recover perturbative, weak coupling behaviour and the non-interacting limit of the problem.

The theoretical difficulties here are well known to be considerable, even for a pure AIM and naturally the more so for self-consistent lattice models; particularly in the context of dynamical properties such as single-particle excitation spectra, optical conductivities and associated transport properties. The PAM itself has of course been studied extensively within DMFT \cite{3,4,5,6}: via numerical methods such as the numerical renormalization group (NRG) \cite{7,8}, quantum Monte Carlo (QMC) \cite{9,10,11,12} and exact diagonalization (ED) \cite{13}; and theoretical approaches that include perturbation theory (PT) in the interaction strength \cite{14,15}, iterated perturbation theory (IPT) \cite{16,17}, the lattice non-crossing approximation (LNCA) \cite{18,19} and the simpler but related average t-matrix approximation \cite{20}, large-$N$ mean-field theory \cite{21,22}, and the Gutzwiller approach \cite{40,44}. While much has been learned from these techniques, most suffer from well recognised limitations and only the NRG \cite{7,8} meets the above criteria. For example, finite-size effects render QMC and ED of limited value, the former 
being further confined to modest interaction strengths and finite temperature ($T$). On the theoretical side, perturbative methods are restricted to weak coupling and cannot recover exponentially small scales; NCA-based approaches violate Fermi liquid behaviour at low-energies and do not recover the non-interacting limit; while large-N mean-field theory, which cannot handle the full range of interaction strengths, amounts simply to a renormalization of the non-interacting limit and as such is applicable only on the very lowest-energy scales.

The above comments underscore the need for the development of new, necessarily approximate theories. In the context of pure quantum impurity models (AIMs) we have recently initiated one such: the local moment approach (LMA) \cite{23,24,25,26,27,28,29,30,31}, whose primary focus is dynamics and transport properties. The LMA satisfies the desiderata listed above. It handles all energy scales on an equal footing, and its intrinsically non-perturbative nature enables it to capture the spin-fluctuation physics characteristic of the strong coupling Kondo regime, embodied dynamically in the exponentially narrow, low-energy Kondo resonance; yet it also spans the full range of interactions, reducing asymptotically in weak coupling to straight second order PT in $U$ \cite{23,28,29}. Fermi liquid behaviour is moreover recovered at low energies - where appropriate, the latter point emphasising that the approach is not confined to the conventional Fermi liquid physics inherent to the metallic AIM, but can also handle the non-Fermi liquid behaviour and associated quantum phase transition arising e.g.\ in the pseudogap AIM \cite{29,30,31}. Symmetric \cite{23,24,25,26,27,29,30,31} and asymmetric \cite{28} AIMs can now be handled within the LMA, which has also been extended to incorporate finite-$T$ \cite{27} and the role of an applied magnetic field $H$ \cite{25,26}. Results for dynamics arising therefrom have also been shown \cite{24,26,27,28,29,30} to give very good agreement with NRG calculations; and, for static magnetic properties of the metallic AIM, with exact results provided by the Bethe ansatz \cite{25,26}.

In this paper we develop the LMA within DMFT to encompass the symmetric PAM \cite{7,8,9,10,13,14,16,19}, well known to be important in understanding the class of mainly cubic Kondo insulating materials such as CeNiSn, $\mathrm{SmB}_6$, $\mathrm{Ce}_3\mathrm{Bi}_4\mathrm{Pt}_3$ and $\mathrm{YbB}_{12}$ (see e.g.\ \cite{32,33}), whose low-energy electronic structure is characterised by an interaction-ren\-ormalized hybridization gap. Specifically we consider here $T=0$ single-particle dynamics of the paramagnetic phase. After the relevant background (\S\ref{sec:background}), the LMA for the lattice model is specified (\S\ref{sec:LMA}), centering on the two-self-energy description and notion of symmetry restoration (\S\ref{sec:SR}) that underlie the approach. That discussion is general, applicable to an arbitrary diagrammatic approximation for the associated dynamical self-energies, $\Sigma_\sigma(\omega)$; the particular non-perturbative class of diagrams we implement here in practice is specified in \S\ref{sec:self_energies}. Following a brief discussion highlighting deficiencies of the static mean-field approximation (\S\ref{sec:MF}), LMA results for single-particle dynamics are given in \S\ref{sec:results}. As for the pure AIMs considered hitherto \cite{23,24,25,26,27,28,29,30,31}, the LMA passes the criteria outlined above. Our primary emphasis, albeit not exclusive, is naturally on the strong coupling regime of the PAM. Here, granted the ability to capture exponentially small scales characteristic of the Kondo lattice, the resultant clean separation of energy scales enables extraction of the universal scaling behavior of dynamics in terms of the gap scale itself; a successful description of which is in addition a necessary prerequisite for a theory of dynamics and associated transport properties at finite-$T$, which will be considered in a subsequent paper. A brief non-technical summary and some concluding comments are given in \S\ref{sec:conclusion}.
\vspace{-3mm}
\section{Background}
\label{sec:background}
The Hamiltonian for the PAM is given in standard notation by
\begin{eqnarray}
\hat{H}& = & -t\sum_{(i,j),\sigma}c_{i\sigma}^\dagger c^{\phantom\dagger}_{j\sigma}+\sum_{i,\sigma}(\epsilon_f+\tfrac{U}{2}f_{i-\sigma}^\dagger f^{\phantom\dagger}_{i-\sigma})f_{i\sigma}^\dagger f_{i\sigma}^{\phantom\dagger} \nonumber \\
& + & V\sum_{i, \sigma}(f_{i\sigma}^\dagger c_{i\sigma}^{\phantom\dagger} +\mbox{h.c.})
\label{eq:hamiltonian}
\end{eqnarray}
where the first term describes the uncorrelated conduction (c) band with nearest neighbour hopping $t_{ij}=t$. The second refers to the f- levels with site energies $\epsilon_f$ and on-site repulsion $U$, while the final term describes c/f- level hybridization via the local matrix element $V$. We focus here on local single-particle dynamics embodied in $G_{ii;\sigma}^f(\omega) \leftrightarrow G_{ii;\sigma}^f(t)=-i<\hat{T}(f_{i\sigma}^{\phantom\dagger}(t)f_{i\sigma}^\dagger)>$ (and likewise $G_{ii;\sigma}^c(\omega)$ for the c-electrons); and hence the local spectra $D_{ii;\sigma}^\gamma (\omega)=-\pi^{-1}\mbox{sgn}(\omega)\mbox{Im}G_{ii;\sigma}^\gamma(\omega)$ ($\gamma=$ c or f). The key feature of DMFT \cite{3,4,5,6} is that the (f-electron) self-energy is site-diagonal, $\Sigma_{ij;\sigma}^f(\omega)=\delta_{ij}\Sigma_{i\sigma}^f(\omega)$; and from straightforward application of Feenberg's renormalized perturbation theory \cite{34,35}, the $G_{ii;\sigma}^\gamma(\omega)$ are given by
\begin{subequations}
\label{eq:G_general}
\begin{equation}
G_{ii;\sigma}^c(\omega)=\left[\omega^+-\frac{V^2}{\omega^+-\epsilon_f-\Sigma_{i\sigma}^f(\omega)}-S_{i\sigma}(\omega)\right]^{-1}
\label{eq:Gc_general}
\end{equation}
\begin{equation}
G_{ii;\sigma}^f(\omega)=\left[\omega^+-\epsilon_f-\Sigma_{i\sigma}^f(\omega)-\frac{V^2}{\omega^+-S_{i\sigma}(\omega)}\right]^{-1}
\label{eq:Gf_general}
\end{equation}
\end{subequations}
where $\omega^+=\omega+i\mspace{2mu}0^+\mbox{sgn}(\omega)$ and $S_{i\sigma}(\omega) \equiv S_{i\sigma}[\{G_{jj;\sigma}^c\}]$ is the Feenberg (or `medium') self-energy. Equation~(\ref{eq:Gf_general}) embodies the connection to a self-consistent impurity model that is inherent to DMFT \cite{3,4,5,6}, since it may be cast in the `single-impurity' form $G_{ii;\sigma}^f(\omega)=[\omega^+-\epsilon_f-\Sigma_{i\sigma}^f(\omega)-\Delta_{\mathrm{eff}}(\omega)]^{-1}$ with an effective hybridization $\Delta_{\mathrm{eff}}(\omega)=V^2[\omega^+-S_{i\sigma}(\omega)]^{-1}$ that is to be self-consistently determined. 

In this paper we consider explicitly the symmetric PAM with $\epsilon_f=-\frac{U}{2}$, for which $n_f=\sum_\sigma\langle f_{i\sigma}^\dagger f_{i\sigma}^{\phantom\dagger}\rangle =1$ and $n_c=\sum_\sigma \langle c_{i\sigma}^\dagger c_{i\sigma}^{\phantom\dagger} \rangle =1$ for all $U$. In contrast e.g.\ to the Hubbard model, the problem is thus characterized by {\it{two}} independent dimensionless parameters: $U/t_\ast$ and $V/t_\ast$, where the hopping is scaled as $t=t_\ast/2\sqrt{Z_c}$ (with $Z_c \rightarrow \infty$ the co-ordination number).

Equations~(\ref{eq:G_general}) are general: independent of lattice type and whether/not magnetic ordering arises. Here we consider primarily the Bethe lattice (BL) (because its spectrum is bounded), for which the Feenberg self-energy
\begin{equation}
S_{i\sigma}(\omega)=\sum_{j\neq i}t_{ij}^2G_{jj;\sigma}^c(\omega)\;\;.
\label{eq:feenberg}
\end{equation}
We shall moreover focus on the homogeneous paramagnetic phase (the antiferromagnetically ordered state is readily handled but less interesting). In this case equations~(\ref{eq:G_general}) reduce to
\begin{subequations}
\label{eq:G}
\begin{equation}
G^c(\omega)=\left[\omega^+-\frac{V^2}{\omega^+-\Sigma(\omega)}-\frac{1}{4}t_\ast^2G^c(\omega)\right]^{-1}
\label{eq:Gc}
\end{equation}
\begin{equation}
G^f(\omega)=\left[\omega^+-\Sigma(\omega)-\frac{V^2}{\omega^+-\frac{1}{4}t_\ast^2G^c(\omega)}\right]^{-1}
\label{eq:Gf}
\end{equation}
\end{subequations}
written explicitly for the BL; with the conventional single self-energy $\Sigma(\omega)=\epsilon_f+\Sigma^f(\omega)$ ($=\Sigma^{\mbox{\scriptsize{R}}}(\omega)-i\mspace{2mu}\mbox{sgn}(\omega)\Sigma^{\mbox{\scriptsize{I}}}(\omega)$) defined to exclude the trivial Hartree contribution (of $\frac{U}{2}n_f$, which precisely cancels the bare $\epsilon_f=-\frac{U}{2}$). And the problem is particle-hole symmetric, reflected in 
\begin{equation}
\Sigma(\omega)=-\Sigma(-\omega)\;\;\;\;\;\;\;\; G^\gamma(\omega)=-G^\gamma(-\omega)
\label{eq:phs}
\end{equation}
with $\omega=0$ the Fermi level.

The trivial limits of the model, used below, are two-fold. First, $V=0$ (for any $U$), where we denote the c-electron Green function by $g_0(\omega)$, with spectral density $\rho_0(\omega)$: for the BL $\rho_0(\omega)=\frac{2}{\pi t_\ast}[1-(\omega/t_\ast)^2]^{\frac{1}{2}}$ is a semi-ellipse of halfwidth $t_\ast$ (from equation~(\ref{eq:Gc})), while $\rho_0(\omega)=[\sqrt{\pi}t_\ast]^{-1}\mbox{exp}(-[\omega/t_\ast]^2)$ for the hypercubic lattice (HCL) \cite{3,4,5,6}. Second, the non-interacting limit $U=0$, denoting the Green functions by $g_0^\gamma(\omega;V^2)$ with the $V$- dependence explicit ($g_0(\omega)\equiv g_0^c(\omega; V^2=0)$). The corresponding spectra are related generally (from equations~(\ref{eq:G_general})) by
\begin{equation}
d_0^f(\omega; V^2)=\frac{V^2}{\omega^2}d_0^c(\omega; V^2)
\label{eq:U0_dcdf_relate}
\end{equation}
and $d_0^c(\omega;V^2)=\rho_0(\omega-V^2/\omega)$. The obvious point here is that for all $V \neq 0$ the system is a hybridization gap insulator \cite{36}. The gap is soft for the HCL, but hard for the BL where the (half) band-gap $\Delta_g^0(V^2)$ is given by
\begin{equation}
2\Delta_g^0(V^2)=\sqrt{t_\ast^2+4V^2}-t_\ast
\label{eq:U0_gap}
\end{equation}
with corresponding spectrum
\begin{equation}
t_\ast d_0^c(\omega; V^2)=\frac{2}{\pi}\left[1-\left(\frac{V^2}{\omega t_\ast}-\frac{\omega}{t_\ast}\right)^2\right]^{\frac{1}{2}}
\label{eq:U0_dc}
\end{equation}
for $\Delta_g^0 \leq |\omega| \leq \Delta_g^0 +t_\ast$. Note trivially that the $U=0$ spectra cannot be expressed in a one-parameter scaling form by suitable dimensionless rescaling of $\omega$: no matter how $\omega$ is thus rescaled (e.g.\ as $\omega/t_\ast$), the $d_0^\gamma$ each remain dependent on the ratio $V/t_\ast$ of bare parameters.

On increasing $U$ from zero the system remains insulating and is perturbatively connected to the non-interacting limit, being as such a Fermi liquid (which is wholly compatible with the insulating nature of the state). The limiting low-$\omega$ behaviour of the single-particle Green functions amounts to a renormalization of the non-interacting limit, which is the origin of the renormalized band picture \cite{2,5,37}. This follows simply by employing the leading low-$\omega$ expansion of the self-energy $\Sigma(\omega)$ \cite{2,5}, viz
\begin{equation}
\Sigma(\omega) \sim - \left[\frac{1}{Z}-1\right]\omega
\label{eq:low_w_SE}
\end{equation}
from equation~(\ref{eq:phs}), with $Z=[1-(\partial \Sigma^{\mbox{\scriptsize{R}}}(\omega)/\partial \omega)_{\omega=0}]^{-1}$ the quasiparticle weight (and with $\Sigma^{\mbox{\scriptsize{I}}}(\omega)$ neglected on the grounds that it vanishes in the gap). The leading low-$\omega$ behaviour of the $G^\gamma(\omega)$ then follows (from equations~(\ref{eq:G_general})) as:
\begin{subequations}
\label{eq:low_w_G}
\begin{eqnarray}  
 G^c(\omega) & \sim & g_0^c(\omega; ZV^2)
\label{eq:low_w_Gc} \\
G^f(\omega) & \sim & Zg_0^f(\omega; ZV^2)
\label{eq:low_w_Gf}
\end{eqnarray}
\end{subequations}

Equations~(\ref{eq:low_w_G}) embody the quasiparticle behaviour of the PAM, akin to the local Fermi liquid quasiparticle form for the impurity Green function of the Anderson impurity model (AIM) \cite{2}. While well known {\it per se} \cite{2}, they have an important implication for the scaling behaviour of the single-particle spectra $D^\gamma(\omega)$ in the strong coupling/Kondo lattice regime of large-$U$, where the quasiparticle weight $Z$ becomes exponentially small \cite{8} (as considered in \S~\ref{sec:results}). The renormalized indirect band-gap $\Delta_g=\Delta_g^0(ZV^2)$ is given from equation~(\ref{eq:U0_gap}) as $Z \rightarrow 0$ by $\Delta_g=ZV^2/t_\ast$, and the spectra $D^\gamma(\omega)$ follow from equations~(\ref{eq:low_w_G}, \ref{eq:U0_dc}, \ref{eq:U0_dcdf_relate}). Their scaling behaviour in strong coupling follows by considering finite $\omega^{\prime}=\omega/\Delta_g$ in the formal limit $\Delta_g \propto Z \rightarrow 0$; and the resultant asymptotic behaviour of the $D^\gamma(\omega)$ is given by
\begin{subequations}
\label{eq:renormU0}
\begin{eqnarray}
t_\ast D^c(\omega) & \sim & \frac{2}{\pi}\left[1-\frac{1}{\omega^{\prime 2}}\right]^{\frac{1}{2}} \\
\label{eq:Dc_renormU0}
2\pi\frac{V^2}{t_\ast}D^f(\omega) & \sim &\frac{4}{\omega^{\prime 2}}\left[1-\frac{1}{\omega^{\prime 2}}\right]^{\frac{1}{2}}
\label{eq:Df_renormU0}
\end{eqnarray}
\end{subequations}

Equations~(\ref{eq:renormU0}) show that both $t_\ast D^c(\omega)$ and $\frac{V^2}{t_\ast}D^f(\omega)$ (and {\it not} therefore $t_\ast D^f(\omega)$) exhibit one-parameter universal scaling in terms of $\omega^{\prime}=\omega/\Delta_g$, with {\it no} explicit dependence on the bare material parameters $U/t_\ast$ and $V/t_\ast$ (which behaviour is naturally not specific to the BL). This simple argument does not of course determine either the dependence of $\Delta_g$ on the bare parameters, or more importantly the $\omega^{\prime}$- range in which equations~(\ref{eq:low_w_G}, \ref{eq:renormU0}) hold: for these a `real' theory is required. But the observation is important in that it suggests the $D^\gamma(\omega)$ in strong coupling should more generally exhibit such scaling, and provides explicitly the limiting behaviour that, as $|\omega^{\prime}|=|\omega|/\Delta_g \rightarrow 0$, must of necessity be recovered by any microscopic theory; which touchstone will be compared to the LMA results in section~\ref{sec:LMAI_II} .

\section{Local Moment Approach (LMA)}
\label{sec:LMA}
The usual route to single-particle dynamics is via the conventional single self-energy $\Sigma(\omega)$. But a determination of $G(\omega)$ in this way is neither mandatory nor a priori desirable: in practice, theoretical approaches of this ilk are liable to suffer from being essentially perturbative in the interaction $U$ (even if self-consistent), thus preventing access to the strongly correlated regime of primary interest. The LMA thus avoids such an approach completely and, as for the pure impurity models considered hitherto \cite{23,24,25,26,27,28,29,30,31}, has three essential elements. (i) Local moments (`$\mu$'), regarded as the first effect of interactions, are introduced explicitly and self-consistently from the outset: the starting point is thus static mean-field (MF, i.e.\ unrestricted Hartree-Fock). This contains two {\it degenerate, local} symmetry broken MF states, denoted by $\alpha=$ A or B and corresponding respectively to local moments $\mu=+|\mu|$ or $-|\mu|$. Severely limited by itself, simple MF nonetheless provides a starting point for a non-perturbative many-body approach. (ii) The LMA achieves this by employing a two-self-energy description that follows naturally from the underlying two local saddle points; introducing non-trivial dynamics into the two self-energies via their functional dependence on the broken symmetry MF propagators. (iii) The final, key idea behind the LMA is that of symmetry restoration: self-consistent restoration of the broken symmetry inherent at pure MF level, and recovery of Fermi liquid/quasiparticle behaviour, as discussed below.

As for the paramagnetic phase of the Hubbard model in infinite-d \cite{38}, the essence of the approach to the PAM, whether at MF level or beyond, is statistical: any given site is with equal probability of $\alpha=$A ($\mu=|\mu|$) or B ($\mu=-|\mu|$) type. First consider briefly pure MF, where the interaction self-energies are purely static Fock (`bubble diagram') contributions; given explicitly by $\tilde\Sigma_{\mbox{\scriptsize{A}}\sigma}^0=-\frac{\sigma}{2}U|\mu|=\tilde\Sigma_{\mbox{\scriptsize{B}}-\sigma}^0$ for $\alpha=$ A or B sites. The corresponding MF propagators are denoted by $g_{\alpha \sigma}^{\gamma}(\omega)$ ($\gamma=$ c or f), and the total Green functions by
\begin{equation}
g^{\gamma}(\omega)=\frac{1}{2}\sum_\alpha g_{\alpha \sigma}^\gamma (\omega)\;\;.
\label{eq:MF_average}
\end{equation}
The $g_{\alpha \sigma}^{\gamma}(\omega)$ are given for the BL by (see equations~(\ref{eq:G_general})) 
\begin{subequations}
\label{eq:MF_g}
\begin{equation}
g_{\mbox{\scriptsize{A}}\sigma}^c(\omega)=\left[\omega^+-\frac{V^2}{\omega^+ + \sigma x}-\frac{1}{4}t_\ast^2 g^c(\omega)\right]^{-1}
\label{eq:MF_gc}
\end{equation}
\begin{equation}
g_{\mbox{\scriptsize{A}}\sigma}^f(\omega)=\left[\omega^+ +\sigma x -\frac{V^2}{\omega^+ -\frac{1}{4}t_\ast^2 g^c(\omega)}\right]^{-1}
\label{eq:MF_gf}
\end{equation}
\end{subequations}
where $\x$; and where (see equation~(\ref{eq:feenberg})) the Feenberg self-energy $S_{i\sigma}(\omega)=\frac{1}{4}t_\ast^2g^c(\omega)$ since precisely half the ($Z_c \rightarrow \infty$) nearest neighbours to any given site are of $\alpha=$ A (or B) type. The $g_{\mbox{\scriptsize{B}}\sigma}^{\gamma}(\omega)$ follow analogously and satisfy $g_{\mbox{\scriptsize{A}}\sigma}^{\gamma}(\omega)=g_{\mbox{\scriptsize{B}}-\sigma}^{\gamma}(\omega)$ (`$\uparrow\mspace{-5mu}/\mspace{-5mu}\downarrow$-spin symmetry'); while particle-hole symmetry is reflected in $g_{\alpha \sigma}^{\gamma}(\omega)=-g_{\alpha -\sigma}^{\gamma}(-\omega)$ and hence (via equation~(\ref{eq:MF_average})) $g^{\gamma}(\omega)=-g^{\gamma}(-\omega)$. And at pure MF level the local moment $|\mu|$ is determined self-consistently from the usual MF condition $|\mu|=|\bar\mu(x)|$, with $|\bar\mu|$ given by
\begin{equation}
|\bar\mu|=\int_{-\infty}^0d\omega\;[d_{\mbox{\scriptsize{A}}\uparrow}^f(\omega)-d_{\mbox{\scriptsize{A}}\downarrow}^f(\omega)]
\label{eq:moment}
\end{equation}
(and $d_{\alpha \sigma}^{\gamma}(\omega)=-\pi^{-1}\mbox{sgn}(\omega)\mbox{Im}g_{\alpha\sigma}^{\gamma}(\omega)$).

MF results will be discussed in \S\ref{sec:MF}, and their serious deficiencies highlighted. We consider now the general case, the algebraic structure of which is formally equivalent to that at MF level. The full $G^\gamma(\omega)$ are expressed as
\begin{equation}
G^\gamma(\omega)=\frac{1}{2}\sum_{\alpha}G_{\alpha \sigma}^{\gamma}(\omega)
\label{eq:full_average}
\end{equation}
where (see equations~(\ref{eq:G_general}, \ref{eq:feenberg}))
\begin{subequations}
\label{eq:full_G}
\begin{equation}
G_{\alpha\sigma}^c(\omega)=\left[\omega^+-\frac{V^2}{\omega^+-\tilde\Sigma_{\alpha \sigma}(\omega)}-\frac{1}{4}t_\ast^2G^c(\omega)\right]^{-1}
\label{eq:full_Gc}
\end{equation}
\begin{equation}
G_{\alpha\sigma}^f(\omega)=\left[\omega^+-\tilde\Sigma_{\alpha\sigma}(\omega)-\frac{V^2}{\omega^+-\frac{1}{4}t_\ast^2G^c(\omega)}\right]^{-1}
\label{eq:full_Gf}
\end{equation}
\end{subequations}
and the Feenberg self-energy $S_{i\sigma}(\omega)=\frac{1}{4}t_\ast^2G^c(\omega)$ for the same reason given above. In contrast to static MF level however, the self-energies
\begin{equation}
\tilde\Sigma_{\mbox{\scriptsize{A}}\sigma}(\omega)=\tilde\Sigma_{\mbox{\scriptsize{B}}-\sigma}(\omega)
\label{eq:AB_se}
\end{equation} 
are now dynamical. They may be separated conveniently as $\tilde\Sigma_{\mbox{\scriptsize{A}}\sigma}(\omega)=-\frac{\sigma}{2}U|\bar\mu|+\Sigma_{\mbox{\scriptsize{A}}\sigma}(\omega)$ into (a) the purely static Fock contribution $-\frac{\sigma}{2}U|\bar\mu|$ (that alone is retained at pure MF level); together with (b) the dynamical contribution $\Sigma_{\mbox{\scriptsize{A}}\sigma}(\omega)\equiv \Sigma_{\mbox{\scriptsize{A}}\sigma}[\{g_{\mbox{\scriptsize{A}}\sigma}^f\}]$ that is a functional of the MF propagators, and a suitable, naturally approximate choice for which (\S\ref{sec:self_energies}) determines the extent to which the key physics of the problem is captured in practice. The full $G_{\alpha\sigma}^{\gamma}(\omega)$ likewise satisfy $\uparrow\mspace{-5mu}/\mspace{-5mu}\downarrow$-spin symmetry 
\begin{subequations}
\begin{equation}
G_{\mbox{\scriptsize{A}}\sigma}^{\gamma}(\omega)=G_{\mbox{\scriptsize{B}}-\sigma}^{\gamma}(\omega)
\label{eq:spin_symmetry}
\end{equation}
 (from equations~(\ref{eq:full_average}, \ref{eq:AB_se})); as well as particle-hole symmetry
\begin{equation}
G_{\alpha\sigma}^{\gamma}(\omega)=-G_{\alpha -\sigma}^{\gamma}(-\omega)
\label{eq:full_phs}
\end{equation}
\end{subequations}
(reflecting $\tilde\Sigma_{\alpha\sigma}(\omega)=-\tilde\Sigma_{\alpha -\sigma}(-\omega)$ for the symmetric PAM considered). As for the AIMs considered previously \cite{23,24,25,26,27,28,29,30,31}, the two-self-energy description inherent to the LMA is explicit in equations~(\ref{eq:full_average}, \ref{eq:full_G}). The resultant c- and f- electron Green functions are, as they must be in the absence of an applied magnetic field, rotationally invariant: Equation~(\ref{eq:full_average}) is correctly $\sigma$- independent, as follows from equation~(\ref{eq:spin_symmetry}). The requisite particle-hole symmetry $G^{\gamma}(\omega)=-G^{\gamma}(-\omega)$ is likewise satisfied, using equations~(\ref{eq:full_average}, \ref{eq:full_phs}).

Before proceeding note that, using equation~(\ref{eq:spin_symmetry}), equation~(\ref{eq:full_average}) for the $G^\gamma(\omega)$ may be written as $G^{\gamma}(\omega)=\frac{1}{2}\sum_\sigma G_{\alpha \sigma}^{\gamma}(\omega)$, involving now a {\it spin} sum that (from equation~(\ref{eq:spin_symmetry})) is {\it independent} of $\alpha=$ A or B (which obviously holds also for the MF Eq.~(\ref{eq:MF_average})). This is entirely equivalent to equation~(\ref{eq:full_average}). We choose to work with this form and, since the $\alpha$- label is redundant, drop it from now on (using $\alpha=$ A implicitly). The basic underlying equations on which we subsequently focus are thus 
\begin{subequations}
\label{eq:paramag_G}
\begin{equation}
G^{\gamma}(\omega)=\frac{1}{2}\sum_{\sigma}G_{\sigma}^{\gamma}(\omega)
\label{eq:paramag_average}
\end{equation}
with
\begin{equation}
G_{\sigma}^c(\omega)=\left[\omega^+-\frac{V^2}{\omega^+-\tilde\Sigma_\sigma(\omega)}-\frac{1}{4}t_\ast^2G^c(\omega)\right]^{-1}
\label{eq:paramag_Gc} 
\end{equation}
\begin{equation}
G_\sigma^f(\omega)=\left[\omega^+-\tilde\Sigma_\sigma(\omega)-\frac{V^2}{\omega^+-\frac{1}{4}t_\ast^2G^c(\omega)}\right]^{-1}
\label{eq:paramag_Gf}
\end{equation}
\end{subequations}
and the f-electron self-energies separated as
\begin{equation}
\tilde\Sigma_\sigma(\omega)=-\frac{\sigma}{2}U|\bar\mu|+\Sigma_\sigma(\omega)
\label{eq:self_energy}
\end{equation}
with $|\bar\mu|\equiv |\bar\mu(x)|$ given by equation~(\ref{eq:moment}).

\subsection{$\Sigma(\omega)$ and Symmetry Restoration}
\label{sec:SR}
The conventional single self-energy $\Sigma(\omega)$ may if desired be obtained as a direct byproduct of the two-self-energy description inherent to the LMA (although the converse does not of course hold). $\Sigma(\omega)$ is defined by equation~(\ref{eq:Gf}) for $G^f(\omega)$, direct comparison of which to its two-self-energy counterpart equations~(\ref{eq:paramag_average}, \ref{eq:paramag_Gf}) yields
\begin{equation}
\Sigma(\omega)=\tfrac{1}{2}(\tilde\Sigma_\uparrow(\omega)+\tilde\Sigma_\downarrow(\omega))+\frac{[\frac{1}{2}(\tilde\Sigma_\uparrow(\omega)-\tilde\Sigma_\downarrow(\omega))]^2}{{\cal{G}}^{-1}(\omega)-\frac{1}{2}(\tilde\Sigma_\uparrow(\omega)+\tilde\Sigma_\downarrow(\omega))}
\label{eq:single_se}
\end{equation}
where ${\cal{G}}(\omega)=[(G^f(\omega))^{-1}+\Sigma(\omega)]^{-1}$ is the so-called host/ medium f-electron propagator \cite{39}. This relation is general (i.e. not specific to the BL), but for the particular case of the BL ${\cal{G}}(\omega)$ is given explicitly by
\begin{equation}
{\cal{G}}^{-1}(\omega)=\omega^+-\frac{V^2}{\omega^+-\frac{1}{4}t_\ast^2G^c(\omega)}
\label{eq:scriptg}
\end{equation}
(using equation~(\ref{eq:Gf})). Notice also that $\Sigma(\omega)$ can equivalently be defined by equation~(\ref{eq:Gc}) for $G^c(\omega)$: direct comparison of which to its two-self-energy counterpart equations~(\ref{eq:paramag_average}, \ref{eq:paramag_Gc}) again yields equation~(\ref{eq:single_se}), as it must. Given the $\{\tilde\Sigma_\sigma(\omega)\}$ and hence (from equations~(\ref{eq:paramag_average}, \ref{eq:paramag_Gc})) $G^c(\omega)$, equation~(\ref{eq:single_se}) enables $\Sigma(\omega)$ to be determined.

The final, important notion underlying the LMA is symmetry restoration (SR) \cite{23,24,25,26,27,28,29,30,31}: self-consistent restoration of the broken symmetry endemic at pure MF level, and correct recovery of the low-$\omega$ quasiparticle behaviour equations~(\ref{eq:low_w_G}) that reflects adiabatic continuity to the non-interacting limit. As for the AIMs considered hitherto \cite{23,24,25,26,27,28,29,30,31}, whether symmetric or asymmetric, this is embodied in the SR condition $\tilde\Sigma_\uparrow(\omega=0)=\tilde\Sigma_\downarrow(\omega=0)$ at the Fermi level; and hence $\tilde\Sigma_\sigma(\omega=0)=0$ (for either $\sigma$) for the present particle-hole symmetric problem ($\tilde\Sigma_\sigma(\omega)=-\tilde\Sigma_{-\sigma}(-\omega)$), i.e.\
\begin{equation}
\tilde\Sigma_\uparrow(\omega=0)=\Sigma_{\uparrow}(\omega=0)-\frac{1}{2}U|\bar\mu|=0\;\;.
\label{eq:SR}
\end{equation}
If SR equation~(\ref{eq:SR}) is satisfied, then the leading $\omega \rightarrow 0$ behaviour of the $\tilde\Sigma_\sigma(\omega)$ follows from particle-hole symmetry as $(\mbox{Re}\tilde\Sigma_\sigma(\omega) \equiv) \tilde\Sigma_\sigma^{\mbox{\scriptsize{R}}}(\omega)=-(Z_\sigma^{-1}-1)\omega$, where $Z_\sigma=[1-(\partial \tilde\Sigma_\sigma^{\mbox{\scriptsize{R}}}(\omega)/\partial \omega)_{\omega=0}]^{-1}$ is thus defined and is independent of $\sigma$. (The $\tilde\Sigma_\sigma^{\mbox{\scriptsize{I}}}(\omega)\equiv \Sigma_\sigma^{\mbox{\scriptsize{I}}}(\omega)$ may be neglected inside a hard gap or, if the gap is soft, because they vanish sufficiently rapidly as $\omega \rightarrow 0$; which behaviour is guaranteed from the diagrams for $\Sigma_\sigma(\omega)$, \S\ref{sec:self_energies}ff.) Using this asymptotic behaviour in the basic two-self-energy equation~(\ref{eq:paramag_G}) for the $G^\gamma(\omega)$ shows that the quasiparticle form equation~(\ref{eq:low_w_G}) is correctly recovered, with quasiparticle weight $Z=[1-(\partial \Sigma^{\mbox{\scriptsize{R}}}(\omega)/\partial \omega)_{\omega=0}]^{-1} \equiv Z_\sigma$; i.e.\ the leading low-$\omega$ behaviour of $\tilde\Sigma_\sigma^{\mbox{\scriptsize{R}}}(\omega)$ and $\Sigma^{\mbox{\scriptsize{R}}}(\omega)$ coincide, $\tilde\Sigma_\sigma^{\mbox{\scriptsize{R}}}(\omega)=\Sigma^{\mbox{\scriptsize{R}}}(\omega)=-(Z^{-1}-1)\omega$ (as may also be verified directly from equation~(\ref{eq:single_se})). And we add that the persistence of the insulating gap with increasing interaction $U$, implied by the quasiparticle form equations~(\ref{eq:low_w_G}) with $Z>0$, is readily shown to be guaranteed only if SR equation~(\ref{eq:SR}) is satisfied. Finally, we note that the SR condition $\tilde\Sigma_\uparrow(\omega=0) = \tilde\Sigma_\downarrow(\omega=0)$ can be shown to apply generically to the paramagnetic phase of the asymmetric PAM, which is in general metallic. In this case, as for the asymmetric AIM discussed in \cite{28}, satisfaction of SR ensures that $\Sigma^{\mbox{\scriptsize{I}}}(\omega) \propto \omega^2$ as $\omega \rightarrow 0$ and hence the recovery of metallic Fermi liquid behaviour as required by continuity to the non-interacting limit. Equation~(\ref{eq:SR}) is thus a particular case of the general SR condition, applicable to the particle-hole symmetric PAM where the ground state is insulating. As for the AIM its imposition --- as a single condition {\it at} the Fermi level $\omega=0$ --- underlies the LMA, amounting in practice to a self-consistent determination of the local moment $|\mu|$ (that supplants the pure MF condition $|\mu|=|\bar\mu(x)|$, see equation~(\ref{eq:moment})); and, most importantly, generating the low-energy spin-flip/Kondo scale that is symptomatic of the Kondo insulators and detailed in section~\ref{sec:results}. 

\subsection{Self-energies}  
\label{sec:self_energies}

\begin{figure}
\begin{center}
\includegraphics*[width=88mm]{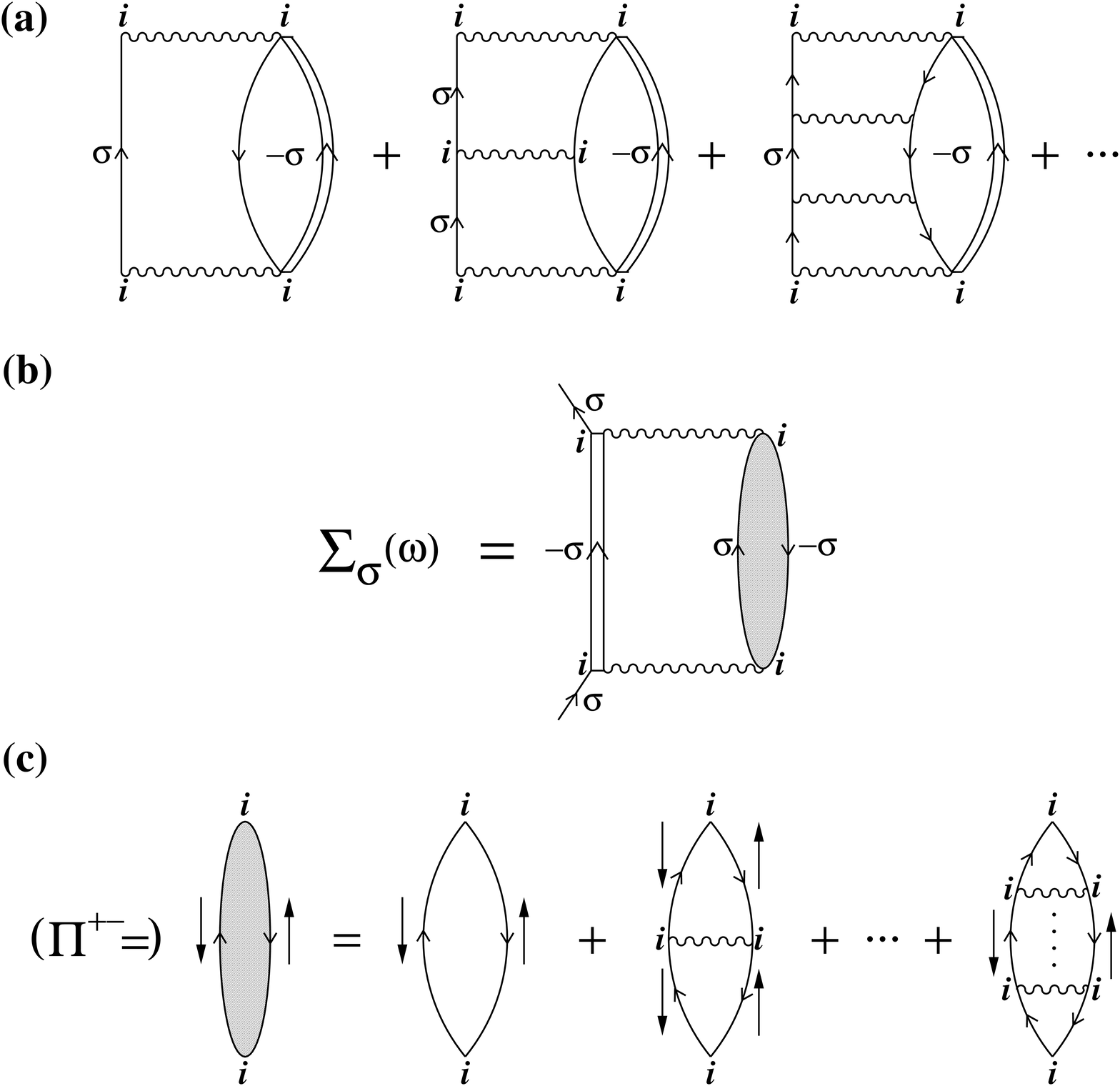}
\end{center}
\caption{(a) Class of diagrams for the f-electron self-energies $\Sigma_\sigma(\omega)$ here retained in practice. The interaction $U$ is denoted by a wavy line, MF propagators by a single line, and the renormalized host/medium propagator (see text) by a double line. (b) Equivalent recasting (including incoming/outgoing propagators) to illustrate the spin-flip scattering involved. (c) Particle-hole ladder sum in the transverse spin channel; for $\Pi^{-+}(\omega)$, spins are reversed.}
\label{fig:feynman}
\end{figure}
The above discussion is general, and to proceed in practice requires specification of the dynamical $\Sigma_\sigma(\omega)$'s (equation~(\ref{eq:self_energy})). The specific class of diagrams retained here is naturally motivated on physical grounds. They embody self-consistent, dynamical coupling of single-particle excitations to low-energy transverse spin fluctuations on all sites, that is essential in particular to capture the strong coupling Kondo lattice regime. The diagrams are shown in Figure~\ref{fig:feynman}, single lines denoting the MF f-electron propagators for site $i$, and wavy lines the local interaction $U$; the double line propagator denotes the broken symmetry host/medium f-electron propagator $\tilde{\cal{G}}_{\sigma}(\omega)$ specified below (see also \cite{38}). The diagrams may be regrouped as shown in Figure~\ref{fig:feynman}, which translates to:
\begin{equation}
\Sigma_\sigma(\omega)=U^2\int_{-\infty}^{\infty}\frac{d\omega_1}{2\pi i}\;\tilde{\cal{G}}_{-\sigma}(\omega-\omega_1)\Pi^{-\sigma \sigma}(\omega_1)
\label{eq:dynamical_se}
\end{equation}

In physical terms these diagrams describe dynamical, correlated spin-flip scattering processes: in which having, say, added a $\sigma$-spin electron to a $-\sigma$-spin occupied f-level on site $i$, the $-\sigma$-spin hops off the f-level and thus generates an on-site spin-flip (reflected in the transverse spin polarization propagator $\Pi^{-\sigma \sigma}(\omega)$); the $-\sigma$-spin electron then propagates through the lattice/host in a correlated fashion, interacting {\it fully} with f-electrons on sites $j \neq i$ (reflected in the host/medium $\tilde{\cal{G}}_{-\sigma}(\omega)$); before returning to site $i$ at a later time whereupon the originally added $\sigma$-spin is removed (which process simultaneously restores the spin-flip on site $i$). The renormalized f-electron medium propagator $\tilde{\cal{G}}_{-\sigma}$, which embodies correlated propagation of the $-\sigma$-spin electron through the lattice, is given explicitly for the BL by (cf. equation~(\ref{eq:scriptg}))
\begin{equation}
\tilde{\cal{G}}_{-\sigma}(\omega)=\left[\omega^+-\frac{\sigma}{2}U|\mu|-\frac{V^2}{\omega^+-\frac{1}{4}t_\ast^2G^c(\omega)}\right]^{-1}\;\;.
\label{eq:scriptg_tilde}
\end{equation}
 Its diagrammatic expansion in terms of MF propagators and dynamical self-energy insertions $\Sigma_{-\sigma}(\omega)$, and hence the infinite set of diagrams implicitly summed in equation~(\ref{eq:dynamical_se}) for $\Sigma_\sigma(\omega)$, is readily shown to have precisely the same topology as its counterpart for the Hubbard model employed previously in \cite{38}. We thus refer the reader to \cite{38} (section 3.1 therein) for detailed discussion of the diagrammatics, here emphasizing just two points. (i) Renormalization of $\Sigma_\sigma(\omega)$ in terms of the medium propagator $\tilde{\cal{G}}_{-\sigma}(\omega)$, rather than $G_{-\sigma}^f$ itself, is embodied in the fact \cite{38} that $\tilde{\cal{G}}_{-\sigma}$, contains dynamic self-energy insertions on any site $j$ {\it excluding}  the original site $i$ (as implicit in the static MF interaction on site $i$ appearing in equation~(\ref{eq:scriptg_tilde})). This accounts in effect for the hard core boson nature of the on-site spin-flip, which would be violated if renormalization in terms of $G_{-\sigma}^f$ was employed (itself necessitating inclusion of additional classes of cancelling diagrams). (ii) Most importantly we emphasize the strongly renormalized, and hence self-consistent, nature of the diagrams retained in $\Sigma_\sigma(\omega)$. This is reflected directly in the fact that $\tilde{\cal{G}}_{-\sigma}(\omega)$ (equation~(\ref{eq:scriptg_tilde})) depends upon $G^c(\omega)$ (via the `effective hybridization' $\Delta_{\mathrm{eff}}(\omega)=V^2[\omega^+-\frac{1}{4}t_\ast^2G^c(\omega)]^{-1}$), which is to be self-consistently determined via solution of equations~(\ref{eq:paramag_G}).

The transverse spin polarization propagator entering equation~(\ref{eq:dynamical_se}) for $\Sigma_\sigma(\omega)$ is given at the simplest level, shown explicitly in Figure~\ref{fig:feynman}, by an RPA-like particle-hole ladder sum in the transverse spin channel; viz
\begin{equation}
\pisigsig(\omega)={}\pisigsigzero(\omega)\left[1-U\mspace{2.5mu}\pisigsigzero(\omega)\right]^{-1}
\label{eq:RPA}
\end{equation}
where the bare particle-hole bubble is itself expressed in terms of the broken symmetry MF f-electron propagators $\{g_\sigma^f\}$. Our subsequent discussion refers explicitly to $\pisigsig(\omega)$ described at this level; in \S\ref{sec:results} a further renormalization is also considered, wherein $\pisigsigzero(\omega)$ and hence $\pisigsig(\omega)$ are expressed in terms of the fully self-consistent host/medium propagators $\{\tilde{\cal{G}}_\sigma\}$ (i.e.\ with {\it{all}} propagators in Figure~\ref{fig:feynman} renormalized in terms of the double line $\tilde{\cal{G}}_\sigma$'s). Whichever level is employed, the $\pisigsigzero$ and hence $\pisigsig$ are readily shown to be related by $\Pi^{-\!\sigma \sigma}(\omega) = \pisigsig(-\omega)$; whence only one such, say $\Pi^{+-}(\omega)$, need be considered explicitly. Using this, and the Hilbert transform for $\Pi^{+-}(\omega)$, equation~(\ref{eq:dynamical_se}) reduces to the following form convenient for later analysis (\S\ref{sec:1_loop_lma})
\begin{eqnarray}
\Sigma_{\uparrow}(\omega)=U^2 \int_{-\infty}^{\infty}\frac{d\omega_1}{\pi} &\mbox{Im}&\Pi^{+-}(\omega_1)[\theta(\omega_1)\tilde{\cal{G}}_{\downarrow}^-(\omega_1+\omega)\nonumber \\
& + & \theta(-\omega_1)\tilde{\cal{G}}_{\downarrow}^+(\omega_1+\omega)]
\label{eq:sigma_integral}
\end{eqnarray}
(with $\Sigma_{\downarrow}(\omega)=-\Sigma_{\uparrow}(-\omega)$); where $\theta(x)$ is the unit step function and $\tilde{\cal{G}}_\sigma=\tilde{\cal{G}}_\sigma^++\tilde{\cal{G}}_\sigma^-$ is separated into the one-sided Hilbert transforms
\begin{equation}
\tilde{\cal{G}}_\sigma^{\pm}(\omega)=\int_{-\infty}^{\infty}d\omega_1\frac{\tilde{D}_\sigma(\omega_1)\theta(\pm \omega_1)}{\omega-\omega_1\pm i0^+}
\label{eq:one_sided_HT}
\end{equation}
with $\tilde{D}_\sigma(\omega)=-\pi^{-1}\mbox{sgn}(\omega)\mbox{Im}\tilde{\cal{G}}_\sigma(\omega)$ the corresponding spectral density. Note also that $\Sigma_\sigma(\omega)$ depends explicitly on $U$ (via the interaction vertices), as well as implicitly on $\x$ via the dependence of the MF propagators thereon (equations~(\ref{eq:MF_average}, \ref{eq:MF_g})). In particular, the symmetry restoration condition equation~(\ref{eq:SR}) has the functional form
\begin{equation}
\Sigma_\uparrow^{\mbox{\scriptsize{R}}}(\omega=0; U, x)=\frac{1}{2}U|\bar\mu(x)|
\label{eq:SR_functional_form}
\end{equation}
exploited below.

\subsection{Practice}
\label{sec:practice}
We now draw together the preceeding discussion and specify what we find to be an efficient practical algorithm to solve the LMA-DMFT equations. This consists of the following basic steps:

\noindent(i)`Startup'. For any given $x=\frac{1}{2}U|\mu|$, equations~(\ref{eq:MF_average}, \ref{eq:MF_g}) are first solved to give the MF propagators $g^\gamma(\omega)$ ($\equiv \frac{1}{2}\sum_\sigma g_{\sigma}^{\gamma}(\omega)$ with $g^\gamma_\sigma(\omega)\equiv g_{\mbox{\scriptsize{A}}\sigma}^\gamma(\omega)$ and $\gamma=$c or f). From this the polarization propagators $\Pi^{\sigma -\sigma}(\omega)$ follow (equation~(\ref{eq:RPA})); and with $G^c(\omega)\simeq g^c(\omega)$, the initial f-electron medium propagator $\tilde{\cal{G}}_{-\sigma}(\omega)$ follows on comparing equations~(\ref{eq:scriptg_tilde}, \ref{eq:MF_gf}) as $\tilde{\cal{G}}_{-\sigma}(\omega)\simeq g_{-\sigma}^f(\omega)$. The startup $\Sigma_\sigma(\omega)=\Sigma_\sigma(\omega; U,x)$ follows directly from equation~(\ref{eq:dynamical_se}).

\noindent(ii) The symmetry restoration condition $\Sigma_{\uparrow}^{\mbox{\scriptsize{R}}}(\omega=0; U,x)=\frac{1}{2}U|\bar\mu(x)|$, is then solved for $U$ given $x=\frac{1}{2}U|\mu|$ (or vice-versa which, while entirely equivalent, is less efficient in practice); the local moment $|\mu|$ follows immediately.

\noindent(iii) With the resultant $\tilde\Sigma_\sigma(\omega)=-\frac{\sigma}{2}U|\bar\mu|+\Sigma_\sigma(\omega)$ (equation~(\ref{eq:self_energy})), equations~(\ref{eq:paramag_average}, \ref{eq:paramag_Gc}) are then solved directly for $G^c(\omega)$ (with $G^f(\omega)$ following in consequence from equations~(\ref{eq:paramag_average}, \ref{eq:paramag_Gf})).

\noindent(iv) The resultant $G^c(\omega)$ is then used in equation~(\ref{eq:scriptg_tilde}) to obtain a new $\tilde{\cal{G}}_{-\sigma}(\omega)$; and hence via equation~(\ref{eq:dynamical_se}) a new $\Sigma_\sigma(\omega)$. Now return to step (ii) and iterate to self-consistency.

We find the above algorithm to be efficient, typically converging after 4-5 iterations and computationally fast on a PC. It is easily generalized to the case (\S\ref{sec:results}) where the $\Pi^{+-}$'s are renormalized in terms of the $\{\tilde{\cal{G}}_\sigma\}$; as well as to encompass the hypercubic lattice, section~\ref{sec:LMAI_II} (or indeed an arbitrary lattice DOS $\rho_0(\omega)$ (see \S\ref{sec:background}) that may be employed in materials modelling applications of DMFT).

Results from pure MF (step(i) alone, with $|\mu|$ determined from $|\mu|=|\bar\mu(x)|$) will be discussed in \S\ref{sec:MF}. We shall also discuss separately (\S\ref{sec:1_loop_lma}) `1-loop' results obtained from a single iterative loop of the above scheme (steps (i)-(iii)): this has the advantage of being analytically tractable, and contains much of the key behaviour found in the full iterative solution (\S\ref{sec:LMAI_II}).

Before proceeding we mention the issue of stability that is important for the PAM, as for the AIMs \cite{23,24,25,26,27,28,29,30,31} within the LMA: the fact that $\repipm(\omega=0)>0$ of necessity (as follows directly from its Hilbert transform). For this to be satisfied, using equation~(\ref{eq:RPA}) (and that $\impizeropm(\omega=0)=0$), $0<U\repizeropm(\omega=0)\leq 1$ is required. An explicit expression for $\repizeropm(0)$ is however readily obtained (see e.g.\ \cite{23,28}), viz
\begin{equation}
U\repizeropm(\omega=0)=\frac{|\bar\mu(x)|}{|\mu|}
\label{eq:stability}
\end{equation}
with $|\bar\mu(x)|$ given by equation~(\ref{eq:moment}) (and $\x$). For stability, $|\mu| \geq |\bar\mu(x)|$ is thus required. The broken symmetry pure MF solutions, for which $|\mu|=|\bar\mu(x)|$ determines the local moment, are thus properly stable (always, which we note would {\it not} be the case if restricted Hartree-Fock, with $|\mu|=0$ enforced {\it a priori}, was employed). But they lie on the `stability border', with $U\repizeropm(\omega=0)=1$. The latter in turn implies, from equation~(\ref{eq:RPA}), that the transverse spin propagator $\Pi^{+-}(\omega)$ contains a pole at $\omega=0$; reflecting physically the fact that the pure MF state is, locally, a degenerate doublet. The latter behaviour is correct for a local moment phase, which for the PAM means the zero-hybridization limit where the f-electrons decouple from the conduction band (which limit we add the LMA recovers exactly). It is not however correct for the Kondo insulating state that is adiabatically connected to the non-interacting (singlet) limit: here by contrast the characteristic energy scale for the local spin-flips is non-zero, and on the order of the Kondo scale that typifies the Kondo insulators. 

The above behaviour is however entirely specific to the pure MF level of self-consistency, i.e.\ arises only if $|\mu|$ is determined by $|\mu|=|\bar\mu(x)|$. And the central point is that within the LMA the local moment is determined from the symmetry restoration condition equation~(\ref{eq:SR_functional_form}) (step (ii) above in which $|\mu| > |\bar\mu(x)|$ is always found). In consequence $\impipm(\omega)$ contains not an $\omega=0$ spin-flip pole,  but rather a resonance centred on a non-zero frequency $\omega_m$. This is the low-energy scale characteristic of the Kondo lattice, its origin within the LMA thus stemming from self-consistent imposition of symmetry restoration; and its physical significance being that it sets the timescale $\tau \sim h/\omega_m$ for restoration of the locally broken symmetry/degeneracy inherent at pure MF level. We add moreover that for $V\neq 0$ symmetry restoration is found to be satisfied for all $U \geq 0$. Its breakdown at finite $U$ would signal an underlying quantum phase transition (such as arises in the pseudogap impurity model, see e.g.\ \cite{29,30,31}); which, in concurrence with general belief, does not therefore arise for the paramagnetic phase of the PAM within DMFT.

\section{Mean-field}
\label{sec:MF}
Our discussion of pure MF is brief: it is fundamentally flawed, lacking even an insulating gap beyond small interaction strengths, as illustrated below. Understanding its deficiencies is however important, for it ultimately underpins a successful description of the Kondo insulating state. Moreover as demonstrated in \S\ref{sec:results}, the primary deficiencies of pure MF arise on energy scales on the order of the insulating gap or multiples thereof. Such scales are naturally of paramount importance, but are nonetheless small --- exponentially so in the strong coupling Kondo lattice regime --- compared to bare (`non-universal') scales on the order of $U$, $t_\ast$ or $V$. This suggests that MF might provide an essentially sound description of dynamics on non-universal energy scales; an issue considered in \S\ref{sec:results}.

\begin{figure*}
\begin{center}
\includegraphics*[scale=0.50]{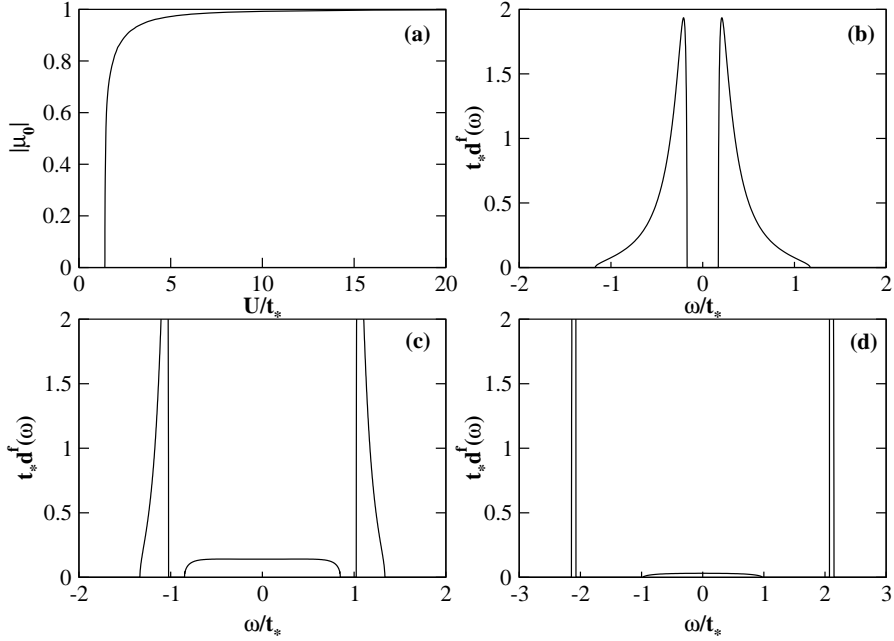}
\end{center}
\protect\caption{Illustrative mean-field results for $V^2/t_\ast^2=0.2$. (a) MF local moment $|\mu_0|$ vs. $U/t_\ast$. MF f-electron spectra, $t_\ast d^f(\omega)$ vs. $\omega/t_\ast$ for (b) $x=\frac{1}{2}U|\mu_0|=0$; (c) $x=0.9 \;(U/t_\ast \simeq 2.1)$; (d) $x=2 \;(U/t_\ast \simeq 4.2)$.}
\label{fig:UHF}
\end{figure*}

Figure~\ref{fig:UHF} shows the $U/t_\ast$ dependence of the MF moment for $V^2/t_\ast^2=0.2$, determined as in \S\ref{sec:LMA} from $|\mu|=|\bar\mu(x)|$ ($\x$) and denoted by $|\mu_0|$. The local moment first becomes non-zero for $U/t_\ast \simeq 1.41$, increasing rapidly thereafter with $U/t_\ast$ such that for $U/t_\ast \gtrsim 2$ or so the moment is well formed and close to saturation. Such behaviour sets in at even lower $U/t_\ast$ with decreasing hybridization $V^2/t_\ast^2$, reflecting the atomic limit incipient as $V\rightarrow 0$ where the f-levels decouple from the band. 

\begin{figure*}
\begin{center}
\includegraphics*[scale=0.50]{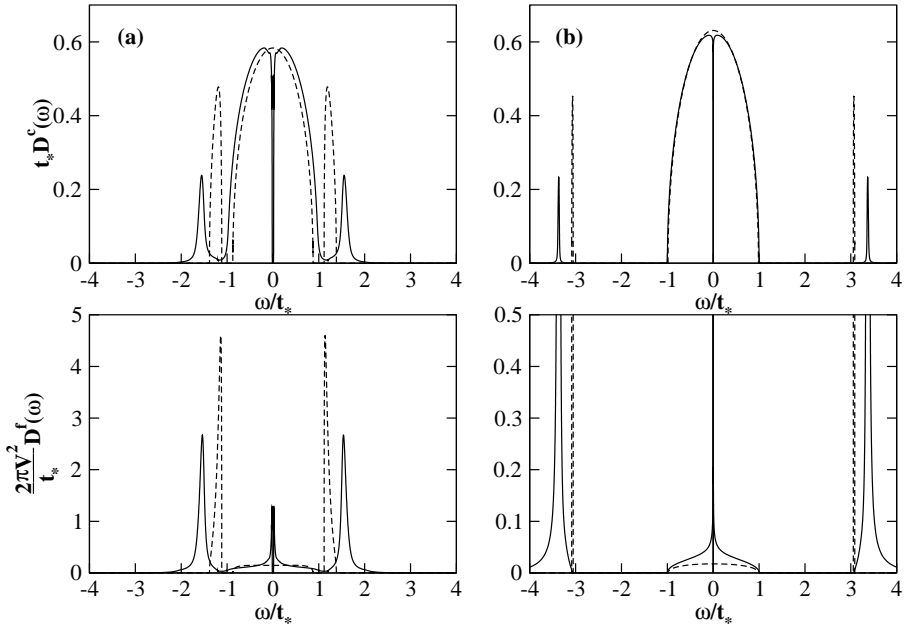}
\end{center}
\caption{LMA spectra, all-scales overview. Top panels: conduction electron spectra; bottom panels: f-electron spectra. (a) For $x=\frac{1}{2}U|\mu| =1\; (U=2.2)$, (b) For $x=3 \;(U=6.1)$; with $V^2=0.2$ in either case. Corresponding results at pure MF level are shown for comparison (dashed lines).}
\label{fig:MF}
\end{figure*}

Illustrative spectral evolution with $U$ is given in Figures~\ref{fig:UHF}(b)-(d), where the f-level spectrum $t_{\ast}d^f(\omega)$ vs. $\omega/t_\ast$ is shown for $x=\frac{1}{2}U|\mu_0|=0, 0.9\, (U/t_\ast \simeq 2.1)$ and $2\; (U/t_\ast \simeq 4.2)$. The characteristics of the non-interacting hybridization gap insulator (Figure~\ref{fig:UHF}(b)) are rapidly lost with increasing $U$: MF produces an insulator-metal transition occurring at $U/t_\ast \simeq 1.48$ just after local moments form (and arising generally at $x=2V^2/t_\ast$ as follows from equations~(\ref{eq:MF_average}, \ref{eq:MF_gc}) at the Fermi level $\omega=0$). With increasing $U/t_\ast$, Figure~\ref{fig:UHF}(c, d), Hubbard satellites at $|\omega| \simeq \frac{U}{2}$ form in the f-electron spectra, rapidly acquiring dominant spectral weight at the expense of intensity on the `band' scales $|\omega| \lesssim t_\ast$; the converse occurring for the MF c-electron spectrum $d^c(\omega)$ (not shown) where spectral intensity on the band scales increases with $U/t_\ast$.

The simple nature of the MF spectra in strong coupling $(U \gg 2V^2/t_\ast)$ is in fact seen directly from equations~(\ref{eq:MF_average}, \ref{eq:MF_g}). For $x=\frac{1}{2}U|\mu_0| \sim \frac{U}{2} \gg V^2/t_\ast$, and for $|\omega| \ll \frac{U}{2}$, equations~(\ref{eq:MF_average}, \ref{eq:MF_gc}) yield 
\begin{equation}
g^c(\omega) \simeq \left[ \omega^+ -\frac{1}{4}t_\ast^2 g^c(\omega)\right]^{-1} \equiv g_0(\omega)
\label{eq:MF_SC_gc}
\end{equation}
 (with $g_0(\omega)$ the $V=0$ conduction band Green function, \S~\ref{sec:background}); i.e.\ $V$ and $U$ drop out, producing a conduction band $d^c(\omega) \simeq \rho_0(\omega)$ characteristic of the decoupled $V=0$ limit (a semi-ellipse for the BL considered explicitly). In practice this behaviour is well attained for $x \gtrsim 2$ or so, and in consequence the f-electron $g_\sigma^f(\omega)$ follows from equation~(\ref{eq:MF_gf}) as $g_{\sigma}^f(\omega)\simeq \left[ \omega^+ + \sigma\frac{U}{2}-V^2 g_0(\omega)\right]^{-1}$. The resultant $d_{\sigma}^f(\omega)$, while dominated by the satellites at $|\omega| \simeq \frac{U}{2}$, nonetheless contains a low-intensity continuum on the band scales $|\omega| < t_\ast$, given explicitly by:
\begin{equation}
d_{\sigma}^f(\omega) \sim 4\frac{V^2}{U^2}\rho_0(\omega)
\label{eq:fband_semi}
\end{equation}
This is evident in Figure~\ref{fig:UHF}; it will also prove important in determining (\S\ref{sec:1_loop_lma}) the $V$-dependence of the Kondo scale within the LMA.

The `insulator-metal transition' is of course entirely an artifact of MF, and the resultant `metal' not a Fermi liquid: the pure MF {\it single} self-energy $\Sigma_{\mbox{\scriptsize{MF}}}(\omega)=\Sigma^{\mbox{\scriptsize{R}}}_{\mbox{\scriptsize{MF}}}(\omega)-i\mbox{sgn}(\omega)\Sigma^{\mbox{\scriptsize{I}}}_{\mbox{\scriptsize{MF}}}(\omega)$ is obtained from the general result equation~(\ref{eq:single_se}) using $\tilde\Sigma_\sigma\equiv -\sigma x$; is given by
\begin{equation}
\Sigma_{\mbox{\scriptsize{MF}}}(\omega)=\frac{x^2}{\omega^+-\frac{V^2}{\omega^+-\frac{1}{4}t_\ast^2g^c(\omega)}}
\label{eq:sigma_MF}
\end{equation}
(which we note is not purely static), and is such that $\Sigma_{\mbox{\scriptsize{MF}}}^{\mbox{\scriptsize{I}}}(\omega=0)\neq 0$ at the Fermi level if $d^c(\omega=0) \propto \mbox{Im}g^c(\omega=0)$ is non-zero --- thus violating Fermi liquid behaviour. But the perversity of MF in predicting a metallic state for the symmetric PAM which describes the Kondo insulator, does not survive the inclusion of correlated electron dynamics in the self-energies $\tilde\Sigma_\sigma(\omega)$. To which we now turn.

\section{Results}
\label{sec:results}
We begin with a brief overview of spectral evolution on all energy scales, and from weak to strong coupling interaction strengths, before turning (\S\ref{sec:1_loop_lma}ff) to the central behaviour at low-energies, and in the strongly correlated regime.

With $t_\ast=1$ as the unit of energy, and $V^2=0.2$, Figure~\ref{fig:MF} shows representative c- and f-electron spectra obtained from the LMA: for $x=\frac{1}{2}U|\mu|=1$ and $3$ (corresponding respectively to $U=2.2$ and $6.1$). Corresponding results at pure MF level are shown for comparison. The $x=1$ example is transitional between weak and strong coupling behaviour. Here the Kondo insulating gap, which the LMA preserves correctly for all $U$, is discernible (just) on the scales shown; and Hubbard satellites, apparent in both $D^f(\omega)$ and $D^c(\omega)$, are just `breaking away' from the main band. The $x=3$ example by contrast is typical of strong coupling behaviour (which in practice is reached by $x \simeq 2$): the gap is exponentially small (\S\ref{sec:1_loop_lma}, \ref{sec:LMAI_II}) and as such not resolved on the scales shown; and the satellites are well formed, dominating the net spectral intensity in $D^f(\omega)$ (as is physically obvious) and giving an ever diminishing contribution to $D^c(\omega)$.

Comparison to the MF spectra is revealing, particularly for the strong coupling example $x=3$ (Figure~\ref{fig:MF}(b)). The Hubbard satellites are, unsurprisingly, broadened and shifted slightly from their MF counterparts. However for the c-electron spectrum it is clear that on the band scales $|\omega| < \tast$, and excepting the (all important) low-energy gap region, the LMA and MF spectra are nigh on coincident; supporting the notion that MF itself provides a reasonable description of the conduction band on bare/non-universal energy scales. We emphasize that this applies only to $D^c(\omega)$ and not to the f-electron spectra: here, as seen directly from Figure~\ref{fig:MF}, the MF spectra are deficient throughout the band, reflecting the dominance of correlated electron dynamics for the f-levels, that are simply absent in MF.
\begin{figure}
\begin{center}
\includegraphics*[scale=0.50]{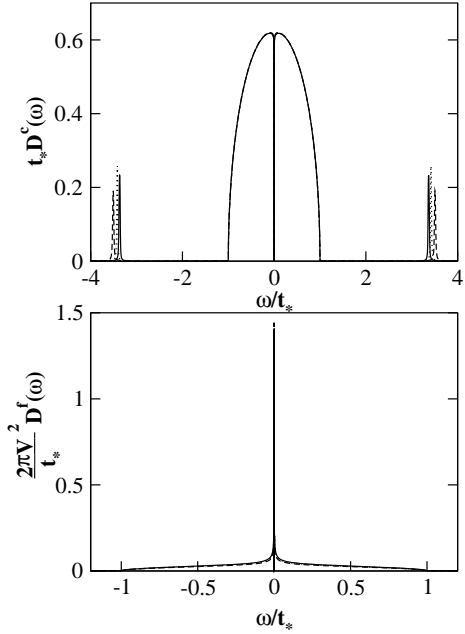}
\end{center}
\caption{All-scales comparison of LMA spectra at different levels of description, for $\x=3 \;(U=6.1)$ with $V^2=0.2$. Top panel, c-electron spectrum; bottom, f-electron spectrum. Solid line: self-consistent LMA (`LMA I', see text, \S\ref{sec:LMAI_II}). Dashed line: with further renormalization of $\pisigsig$ propagators as described in text (`LMA II'). Dotted line: 1-loop LMA results. For the f-electron spectra, the three are barely distinguishable on the scales shown.}
\label{fig:1lma1lma2}
\end{figure}

The LMA results shown in Figure~\ref{fig:MF} refer to the fully self-consistent level described in \S\ref{sec:practice}. We add however that on the full energy regime shown these differ insignificantly from the LMA results obtained either at the simpler 1-loop level (\S\ref{sec:practice}), or with the transverse spin polarization propagators $\Pi^{\sigma -\sigma}$ further renormalized in terms of the host/medium propagators $\{\tilde{\cal{G}}_{\sigma}\}$; as illustrated in Figure~\ref{fig:1lma1lma2} for $x=3$ ($U=6.1$). The differences between these different levels arise of course on the low-energy gap scale, as pursued in the following sections.

\begin{figure}
\begin{center}
\includegraphics*[scale=0.50]{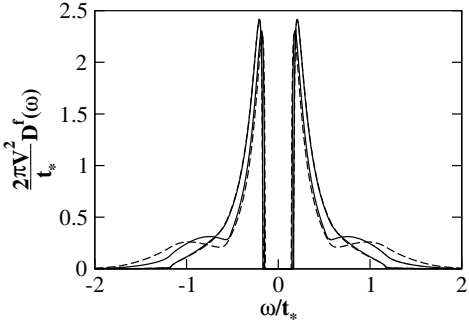}
\end{center}
\caption{Weak coupling LMA f-electron spectra (solid lines), for $U=1$ and $0.25$ (with $V^2=0.2$). Corresponding results from second order perturbation theory in $U$ (SOPT, dashed lines) are also shown; for the lower-$U$ they are indistinguishable from the LMA.}
\label{fig:sopt}
\end{figure}

The comments above refer to intermediate to strong coupling behaviour, with a natural emphasis on the latter. But what of weak coupling? Here, as for the AIMs considered hitherto \cite{23,28,29}, the LMA is readily shown to be perturbatively exact to/including second order in the interaction $U$. Figure~\ref{fig:sopt} demonstrates the point, showing LMA results for $D^f(\omega)$ with $V^2=0.2$ and $U=1$ and $0.25$, compared to those from simple second order perturbation theory in $U$ (SOPT): with decreasing $U$ the LMA spectrum clearly reduces to that from SOPT, and for the lower $U$ shown the two are essentially indistinguishable on all energy scales.

We turn now to the central issues: low-energy dynamics on the scale of the Kondo insulating gap, particularly in the strong coupling regime; and, most importantly, universal spectral scaling in terms of the gap itself.

\subsection{1-Loop LMA}
\label{sec:1_loop_lma}
We begin by discussing the LMA at 1-loop level, viz a single iterative loop (steps (i)-(iii)) of the general scheme discussed in \S\ref{sec:practice}. Results from the full iterative solutions will be considered in \S\ref{sec:LMAI_II}; but the simpler 1-loop level is itself important because it offers an analytical handle on, and insight into, the problem.

\begin{figure*}
\begin{center}
\includegraphics*[scale=0.40]{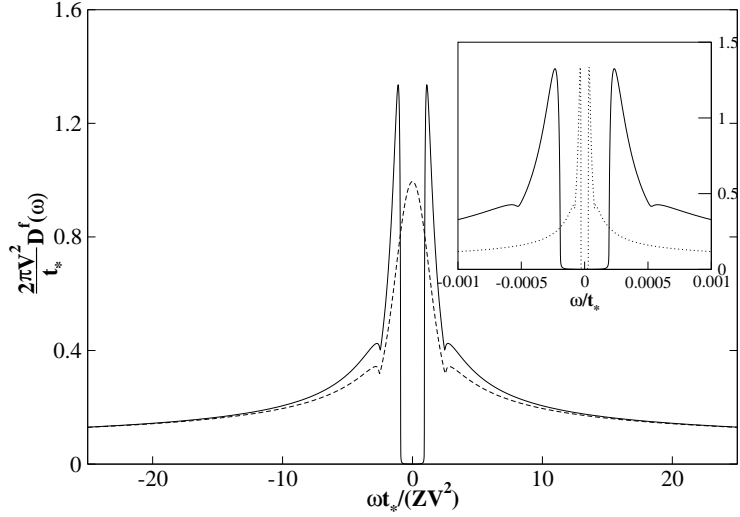}
\end{center}
\caption{1-Loop LMA results for scaling behaviour of f-electron spectrum. Inset: $2\pi\frac{V^2}{t_\ast}D^f(\omega)$ on an `absolute' scale vs. $\omega/t_\ast$, for $U/t_\ast=6.1$ (solid line) and $8.1$ (dotted line). Main figure: collapse to scaling form (solid line), same results shown vs. $\omega t_\ast/(ZV^2) \equiv \omega/\Delta_g$. The scaling spectrum for the Anderson impurity model (see text) is also shown in the main figure (dashed line).}
\label{fig:aimpam1loop}
\end{figure*}

The inset to Figure~\ref{fig:aimpam1loop} shows the low-energy behaviour of the 1-loop f-electron spectrum vs. $\omega (\equiv \omega/t_\ast)$, i.e. on an `absolute' scale; for $V^2=0.2$ and two different interaction strengths, $U=6.1\; (\x=3)$ and $U=8.1\; (x=4)$. The spectra are quite distinct and clearly dependent on the bare material parameters, with the gap narrowing strongly under the modest increase in $U$. The main part of Figure~\ref{fig:aimpam1loop} by contrast shows $2\pi V^2D^f(\omega)$ vs. $\omega^\prime=\omega/[ZV^2]$ where $Z\equiv [1-(\partial \Sigma_\sigma^{\mathrm{R}}(\omega)/\partial \omega)_{\omega=0}]^{-1}$ (\S\ref{sec:self_energies}) is the quasiparticle weight. The point is obvious: the spectra collapse to a common form; which universal scaling is indeed seen to be in terms of the gap scale $\Delta_g=ZV^2$, as required from the quasiparticle form discussed in \S\ref{sec:background} that embodies perturbative continuity to the non-interacting limit (and full comparison to which is given in \S\ref{sec:LMAI_II}, see Figure~\ref{fig:qpf_lma2} below). The scaling spectrum $V^2D^f(\omega)$ is entirely independent of the {\it two} bare parameters $U$ and $V$, which enter solely via the dependence of the gap scale thereon; and we emphasize that this applies to $V^2D^f(\omega)$ and not therefore to $D^f(\omega) [\equiv t_\ast D^f(\omega)]$ (i.e. if the calculations of Figure~\ref{fig:aimpam1loop} are repeated for different $V^2$, $D^f(\omega)$ vs. $\omega^\prime$ is not itself universal but $V^2D^f$ is). Before proceeding we add that, as for the pure impurity models considered hitherto \cite{23,24,25,26,27,28,29,30,31}, the small spectral 'dip' at $|\omega^\prime| \simeq 2.5$ is entirely an artifact of the specific RPA-like form for the $\pisigsig(\omega)$ employed here; it can be eliminated \cite{24} but we are content to live with it in the following.

The first obvious question is how the quasiparticle weight, and hence $\Delta_g=ZV^2$, depends upon the bare parameters in the strong coupling regime of interest. This can be determined analytically as now outlined, focusing on equation~(\ref{eq:sigma_integral}) for $\Sigma_\uparrow^{\mathrm{R}}(\omega)$. At the 1-loop level the medium propagator $\tilde{\cal{G}}_\downarrow$ in equation~(\ref{eq:sigma_integral}) reduces to the MF propagator $g_\downarrow^f$ as explained in section~\ref{sec:practice}, step (i). But in strong coupling the spectral weight of $\impipm(\omega)$ is confined to $\omega>0$, with $\int_0^\infty(d\omega/\pi) \impipm(\omega)=1$; behaviour that reflects physically the strong coupling saturation of the local moment $(|\mu| \rightarrow 1)$. As discussed in \S\ref{sec:practice} the resonance in $\impipm(\omega)$ is (by definition of $\omega_m$) centred on the low-energy spin-flip scale $\omega_m$; and on scales of this order $g_\downarrow^{f-}(\omega)$ is slowly varying, whence equation~(\ref{eq:sigma_integral}) for $\Sigma_\uparrow^{\mathrm{R}}(\omega)$ reduces asymptotically to
\begin{equation}
\Sigma_\uparrow^{\mathrm{R}}(\omega) \sim U^2\mathrm{Re}g_\downarrow^{f-}(\omega+\omega_m)
\label{eq:asymp_sigmaR}
\end{equation}
But $\mathrm{Re}g_\sigma^{f-}(\omega)$ is given by the one-sided Hilbert transform (equation~(\ref{eq:one_sided_HT})); which as $\omega \rightarrow 0$ is dominated by the log singularity arising because (\S\ref{sec:MF}) the MF spectrum $d_\sigma^f(\omega)$ in strong coupling is itself non-vanishing at the Fermi level $\omega=0$. This $\omega \rightarrow 0$ behaviour is captured by
\begin{eqnarray}
\mathrm{Re}g_\downarrow^{f-}(\omega) & \sim & d_\downarrow^f(0)\int_{-t_\ast}^0d\omega_1 \;{\cal{P}}\left(\frac{1}{\omega-\omega_1}\right)\nonumber \\
& = & d_\downarrow^f(0)\,\mathrm{ln}\left(\frac{t_\ast}{|\omega|}\right)
\label{eq:regf_down_minus}
\end{eqnarray}
where a high energy cutoff of order $t_\ast$ is employed (its precise value being immaterial, the important point being that the prefactor to the log divergence is precisely $d^f_\downarrow(0)$). But in strong coupling $d^f_\sigma(\omega=0)$ is given from equation~(\ref{eq:fband_semi}); using which equations~(\ref{eq:asymp_sigmaR}, \ref{eq:regf_down_minus}) yield the desired low-frequency behaviour
\begin{equation}
\Sigma_\uparrow^{\mathrm{R}}(\omega) \sim \;-4V^2 \rho_0(0)\,\mathrm{ln}\left(\frac{|\omega+\omega_m|}{t_\ast}\right)
\label{eq:kondo_scale}
\end{equation}
(with $\rho_0(0)=2/(\pi t_\ast)$ for the BL, although we add that equation~(\ref{eq:kondo_scale}) holds for an arbitrary underlying lattice). From this the quasiparticle weight 

\noindent$\quasi$ follows directly, being given for $\omega_m/V^2\rho_0(0) \rightarrow 0$ by 
\begin{subequations}
\begin{eqnarray}
\frac{\pi}{8}\omega_m & = &\frac{\pi}{2}V^2\rho_0(0)Z 
\label{eq:w_m} \\
& = & \frac{V^2}{t_\ast}Z \equiv \Delta_g
\label{eq:gap}
\end{eqnarray}
\end{subequations}
with the latter explicitly for the BL.

The gap and spin-flip scales are thus equivalent. And their mutual dependence on bare material parameters follows from symmetry restoration (\S\ref{sec:practice}, step (ii)), viz from equation~(\ref{eq:SR}) in strong coupling (where $|\bar\mu|\rightarrow 1$) via $\Sigma_\uparrow^{\mathrm{R}}(\omega=0)=\frac{1}{2}U$; which, combined with equation~(\ref{eq:kondo_scale}) gives the desired result
\begin{subequations}
\label{eq:kondo}
\begin{equation}
\omega_m \sim \Delta_g \sim t_\ast \mathrm{exp}\left(\frac{-U}{8V^2\rho_0(0)}\right)
\label{eq:kondo_scale_general}
\end{equation}
\begin{equation}
=t_\ast \mathrm{exp}\left(\frac{-\pi Ut_\ast}{16V^2}\right)
\label{eq:kondo_scale_pam}
\end{equation}
\end{subequations}
showing that $\Delta_g=ZV^2/t_\ast$ is indeed exponentially small in strong coupling.

Once symmetry has been restored as above, the self-energies $\tilde\Sigma_\sigma(\omega)$ (equation~(\ref{eq:self_energy})) for all $\omega$ follow immediately. The final step in the 1-loop analysis (step (iii), \S\ref{sec:practice}) is to take the resultant $\tilde\Sigma_\sigma(\omega)$ and use them in equations~(\ref{eq:paramag_average}, \ref{eq:paramag_Gc}) to solve self-consistently for $G^c(\omega)$. This in turn determines directly the effective f-electron hybridization $\Delta_{\mathrm{eff}}(\omega)=V^2[\omega^+-S(\omega)]^{-1} \equiv V^2[\omega^+-\frac{1}{4}t_\ast^2G^c(\omega)]^{-1}$ (BL); in terms of which $G^f(\omega)$ and hence $D^f(\omega)$ follow without further ado from equations~(\ref{eq:paramag_average}, \ref{eq:paramag_Gf}), viz
\begin{equation}
G^f(\omega)=\frac{1}{2}\sum_\sigma[\omega^+-\tilde\Sigma_\sigma(\omega)-\Delta_{\mathrm{eff}}(\omega)]^{-1}\;\;.
\label{eq:average_Gf}
\end{equation}
The 1-loop results shown in Figure~\ref{fig:aimpam1loop} have of course been obtained in this way, 

There is however a simpler, certainly crude but nonetheless revealing approximation that may be employed at this final (step (iii)) stage to determine $G^f(\omega)$: namely to replace the $G^c$-dependence of $\Delta(\omega)\equiv \Delta[G^c]$ by $\Delta(\omega) \simeq \Delta[g^c]$ in terms of the MF propagator (as used also in the `startup' step (i) (\S\ref{sec:practice}), there leading to $\tilde{\cal{G}}_{-\sigma} \simeq g_{-\sigma}^f$ in $\Sigma_\sigma(\omega)$). With this, the resultant $\Delta(\omega)=\Delta_{\mathrm{R}}(\omega)-i\mathrm{sgn}(\omega)\Delta_{\mathrm{I}}(\omega)$ reduces in strong coupling, using Eq.~(\ref{eq:MF_SC_gc}), to $\Delta(\omega)\approx V^2g_0(\omega)$ with $g_0(\omega)(=\mathrm{Re}g_0(\omega)-i\mathrm{sgn}(\omega)\pi \rho_0(\omega))$ the $V=0$ conduction electron propagator (a result readily shown to hold for a general lattice). In the strong coupling regime where $\Delta_g \propto Z \rightarrow 0$, the $\omega$-dependence of $g_0(\omega)$ is moreover irrelevant, occurring as it does on non-universal scales on the order of $t_\ast$; whence $\Delta(\omega) \approx -i\mathrm{sgn}(\omega)\pi V^2\rho_0(0)$ and equation~(\ref{eq:average_Gf}) becomes
\begin{equation}
G^f(\omega) \approx \tfrac{1}{2}\sum_\sigma\left[\omega^+-\tilde\Sigma_\sigma(\omega)+i\mathrm{sgn}(\omega)\Delta_0\right]^{-1}
\label{eq:gf_imp}
\end{equation}
where $\Delta_0=\pi V^2\rho_0(0)$. 

Equation~(\ref{eq:gf_imp}) is simply the local impurity Green function for a metallic Anderson impurity model with hybridization strength $\Delta_0=\pi V^2 \rho_0(0) \equiv 2V^2/t_\ast$ (BL). The resultant impurity scaling spectrum $\pi \Delta_0 D_{\mathrm{imp}}(\omega)$ is also compared to the 1-loop $\pi \Delta_0 D^f(\omega)$ in Figure~\ref{fig:aimpam1loop}. The AIM spectrum is of course metallic (with $\pi \Delta_0D_{\mathrm{imp}}(\omega=0)=1$ as required by the Friedel sum rule \cite{2}), in contrast to the gapped PAM case; but beyond the gap the spectral tails of the two rapidly coincide. This is natural, for the nature of the 1-loop calculation is clear: the 1-loop self-energies $\tilde\Sigma_\sigma(\omega)$ are themselves those of the underlying AIM. And since the associated low-energy scale is determined from symmetry restoration via $\tilde\Sigma_\sigma^{\mathrm{R}}(\omega=0)=0$, the low-energy gap scale is simply the Kondo scale for the AIM itself; as seen directly from equation~(\ref{eq:kondo}) expressed as $\Delta_g/t_\ast \sim \mathrm{exp}(-\pi U/8\Delta_0)$, recovering the exact exponent for the symmetric AIM \cite{2}. The gap in the PAM spectrum (which is strictly soft at 1-loop level, albeit not visibly so in Figure~\ref{fig:aimpam1loop}) arises from self-consistent solution of equations~(\ref{eq:paramag_G}a-c) for $G^c(\omega)$ and hence $G^f(\omega)$, {\it given} the effective AIM $\tilde\Sigma_\sigma(\omega)$. In that sense the 1-loop level is akin to an average t-matrix approximation \cite{20}, but formulated within the two-self-energy description inherent to the LMA and with symmetry restoration ensuring Fermi liquid behaviour (in the general sense of adiabatic continuity to the non-interacting limit).

Despite the relative simplicity of the 1-loop LMA, the relation of the AIM implicit therein is both natural and substantially correct: NRG calculations for example \cite{8} (see Figure 1 therein) indeed show, beyond the gap scale, a close similarity between the PAM $D^f(\omega)$ and the AIM spectra. Moreover since the spectral `tails' of the 1-loop PAM and AIM are common (Figure~\ref{fig:aimpam1loop}), recent LMA results for the latter \cite{24} may be used directly to infer their analytical form for $V^2D^f(\omega)$ at 1-loop level; specifically
\begin{eqnarray} 
(\pi V)^2\rho_0(0)D^f(\omega) & \sim &  \frac{1}{2}\left\{\frac{1}{[(4/\pi)\mathrm{ln}(|\tilde\omega|)]^2+1}\right. \nonumber \\
& &+\left.\frac{5}{[(4/\pi)\mathrm{ln}(|\tilde\omega|)]^2+25}\right\}
\label{eq:log_tails}
\end{eqnarray}
where $\tilde\omega=\omega/\omega_m$ (and which form is known \cite{24} to agree quantitatively for $|\tilde\omega| \gtrsim 5$ with NRG results \cite{30} for the AIM itself). These slow logarithmic tails persist in the $V^2D^f(\omega)$ scaling spectrum beyond 1-loop level (\S\ref{sec:LMAI_II}) and, as for the AIM \cite{24,27}, are ultimately responsible for the `high' temperature logarithmic behaviour of transport properties such as the resistivity.

In physical terms the limitations of 1-loop level are nonetheless self evident; it misses the dynamical intersite correlations embodied (as discussed in \S~\ref{sec:self_energies}) in the functional dependence of the self-energies $\Sigma_\sigma(\omega)$ upon the renormalized f-electron medium propagators $\{\tilde{\cal{G}}_\sigma\}$ (which at 1-loop level are simply replaced by their MF counterparts).

\subsection{Fully self-consistent LMA}
\label{sec:LMAI_II}
The full LMA solutions are now considered. In the following, we denote by `LMA I' results obtained from the class of diagrams contributing to $\Sigma_\sigma(\omega)$ shown explicitly in Figure~\ref{fig:feynman}; while `LMA II' refers to those obtained from the further renormalization of $\pisigsig$ in terms of the f-electron medium propagators (i.e.\ {\it{all}} propagators in Figure~\ref{fig:MF} renormalized in terms of the double line $\tilde{\cal{G}}_\sigma$'s). 
\begin{figure}
\begin{center}
\includegraphics*[width=55mm]{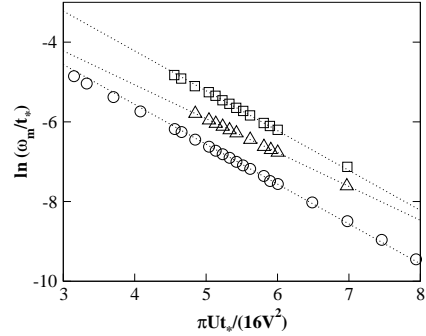}
\end{center}
\caption{Dependence on bare parameters of the low-energy scale $\omega_m \propto \Delta_g = ZV^2/t_\ast$, arising from the LMA at 1-loop (circles), LMA I (squares) and LMA II (triangles).}
\label{fig:kondo_scale}
\end{figure}

\begin{figure*}
\begin{center}{
\includegraphics*[scale=0.40]{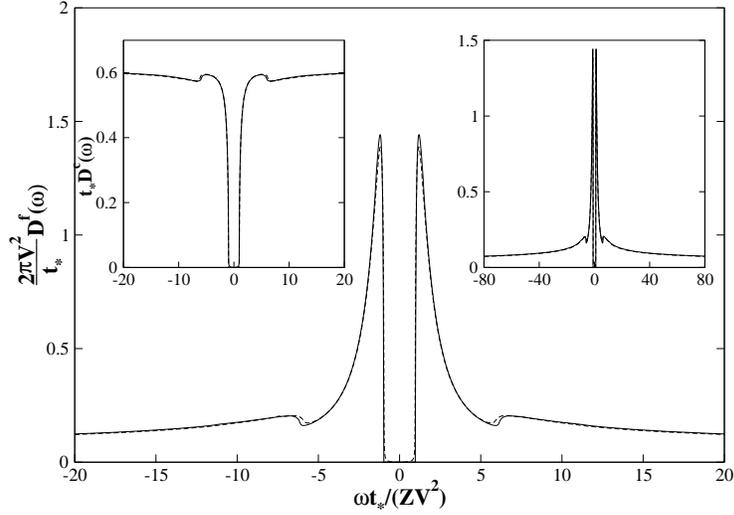}}
\end{center}
\protect\caption{f-electron scaling spectra $2\pi\frac{V^2}{t_\ast}D^f(\omega)$ vs. $\omega/\Delta_g$ arising from LMA I (dashed line) and LMA II (solid line). Right inset: on an expanded scale to show the logarithmic tails. Left inset: corresponding c-electron spectra from LMA I/II.}
\label{fig:scaling}
\end{figure*}

Figure~\ref{fig:kondo_scale} shows the dependence on bare parameters of the low energy scale $\omega_m \propto \Delta_g=ZV^2/t_\ast$ arising from LMA I/II for the Bethe lattice, in comparison to that obtained at 1-loop level (Eq.~(\ref{eq:kondo}), which form is confirmed in Figure~\ref{fig:kondo_scale}, with the prefactor $\sim 0.2t_\ast$ determined numerically). Two points should be noted here. First that the lattice scale is enhanced over its 1-loop counterpart (which as above is equivalently that of the underlying AIM); qualitative behaviour that is also found in NRG \cite{8} and QMC \cite{10} calculations for the symmetric PAM. The second point is the exponential dependence of the scale, $\omega_m \propto \mathrm{exp}[-\lambda U/8V^2\rho_0(0)]$ with $\lambda=1$ at 1-loop level. From Figure~\ref{fig:kondo_scale} we find that $\lambda=1$ remains at the level of LMA I, while $\lambda \simeq 0.85$ is found from the fully renormalized LMA II. The latter is in qualitative agreement with, but somewhat larger than, NRG calculations \cite{8} which yield $\lambda \simeq 0.7$; which in turn are comparably in excess of $\lambda=\frac{1}{2}$ arising from the Gutzwiller approximation \cite{40}. The dependence of the low-energy scale on bare model parameters is however a subsidiary issue compared to the scaling behaviour of dynamics in terms of the gap scale itself. The latter is illustrated in Figure~\ref{fig:scaling} which shows the f-electron scaling spectra $2\pi\frac{V^2}{t_\ast}D^f(\omega)$ vs. $\omega^\prime=\omega/\Delta_g$, for both LMA I and II; the corresponding c-electron spectra $t_\ast D^c(\omega)$ are shown in the left inset. Despite the differences in the absolute values of the low-energy scale (embodied in Figure~\ref{fig:kondo_scale}), the resultant LMA I/II scaling spectra differ very little from each other; and exhibit the same qualitative behaviour as found at 1-loop level (Figure~\ref{fig:aimpam1loop}), including the logarithmic spectral tails in $V^2D^f(\omega)$ evident in the right inset to Figure~\ref{fig:scaling}.

\begin{figure}
\begin{center}
{
\includegraphics*[scale=0.50]{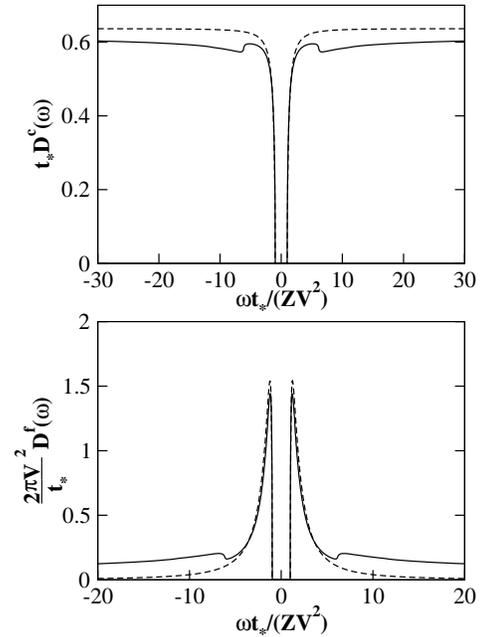}}
\end{center}
\protect\caption{Comparison of LMA II scaling spectra (solid lines) to the limiting low-energy quasiparticle form (dashed lines). Top panel, c-electron scaling spectra; bottom, f-electron spectra.}
\label{fig:qpf_lma2}
\end{figure}

As discussed in \S\ref{sec:background}, adiabatic continuity to the non-interacting limit requires that on sufficiently low energy scales the resultant spectra should conform to the dictates of Fermi liquid theory, as reflected in the quasiparticle form equations~(\ref{eq:low_w_G}, \ref{eq:renormU0}). In Figure~\ref{fig:qpf_lma2} the f- and c-electron scaling spectra arising from LMA II are compared directly to the limiting quasiparticle behaviour equation~(\ref{eq:renormU0})  (corresponding comparisons for LMA I and the 1-loop results are very similar, as evident from Figures~\ref{fig:aimpam1loop},\ref{fig:scaling}). Agreement with the quasiparticle form is essentially perfect close to the gap edges; and in practice is followed quite closely up to $|\omega^\prime|=|\omega|/\Delta_g \sim 3$ or so. Beyond this however, the quasiparticle form fails to capture the logarithmic tails of the scaling spectrum, decaying instead as $V^2D^f(\omega) \sim 1/|\omega^\prime|^2$ (equation~(\ref{eq:Df_renormU0})). This is of course natural since the quasiparticle behaviour is confined strictly to the limiting low-$\omega^\prime$ behaviour, and it mirrors the situation arising in the metallic AIM where the quasiparticle form is a trivial Lorentzian \cite{2} that likewise fails to capture the slow spectral tails of the Kondo resonance \cite{24}.

\begin{figure}
\begin{center}
\includegraphics*[scale=0.50]{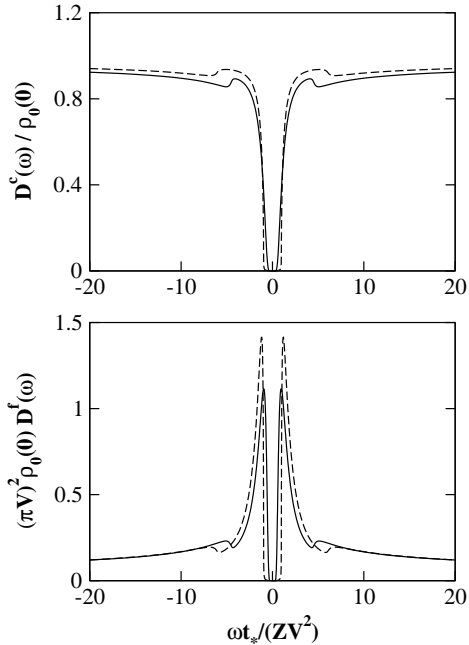}
\end{center}
\caption{LMA I scaling spectra for the hypercubic lattice (solid lines), compared to those for the Bethe lattice (dashed).}
\label{fig:hcl_vs_bl}
\end{figure}

Finally, an obvious question arises: how strongly do the scaling spectra depend upon the host lattice? For the metallic AIM the answer is not at all: in the strong coupling Kondo regime the $\omega$-dependence of the host density of states to which the impurity is coupled is entirely irrelevant (see e.g.\ \cite{2}), and the scaling form of $\pi\Delta_0 D_{\mathrm{imp}}(\omega)$ is host-independent. But this is not of course the case for the PAM. The preceeding results have been given explicitly for the BL where the free ($V=0$) conduction band $\rho_0(\omega)$ is semielliptic. To illustrate the influence of the lattice, Figure~\ref{fig:hcl_vs_bl} compares the LMA I scaling spectra for the BL with those for the hypercubic lattice (HCL, where $\rho_0(\omega)$ is an unbounded Gaussian). While the Kondo insulating gap for the BL is hard, it is strictly soft for the HCL, albeit that this is barely visible in Figure~\ref{fig:hcl_vs_bl} since the resultant $D^\gamma(\omega)$ are exponentially small close to the Fermi level. But beyond the immediate vicinity of the gap, and despite the very different nature of the host $\rho_0(\omega)$'s, the scaling spectra are seen to be qualitatively very similar; supporting the view that for local dynamical properties the one-electron `details' of the lattice play but a minor role.

\section{Discussion/Outlook}
\label{sec:conclusion}
We have developed in this paper a local moment approach to $T=0$ single-particle dynamics of the symmetric PAM, the basic microscopic model for understanding Kondo insulating materials \cite{32,33}. The necessary criteria for a successful description of the problem appear to be met by the LMA, handling as it does all energy scales and interaction strengths while satisfying the requirements of Fermi liquid theory at sufficiently low energies.

Particular attention has been given for obvious physical reasons to the strong coupling (i.e.\ large $U$) Kondo lattice regime, believed to be appropriate to the Kondo insulator materials mentioned in \S\ref{sec:intro}. We have shown that within DMFT+LMA one nevertheless recovers correctly, on sufficiently low energy scales, an `insulating Fermi liquid' behaviour which evolves continuously from the non-interacting hybridization gap insulator. For example, the scaled conduction electron and f-electron spectra have a renormalized non-interacting form (Eqs.~(\ref{eq:low_w_G}, \ref{eq:renormU0})) for $|\omega|/\Delta_g \lesssim 2-3$ (see Fig.~\ref{fig:qpf_lma2}); where $\Delta_g$, the renormalized gap, is reduced from the non-interacting hybridization gap $\Delta_g^0$ by the quasiparticle weight factor $Z$. For larger energies the (scaled) spectra deviate rapidly and substantially from the non-interacting form, but nonetheless remain characterized by a {\it{single}} low-energy scale $\Delta_g \propto Z$ that is exponentially small but in general enhanced over its counterpart in the dilute (AIM) limit \cite{41}. The spectra thus exhibit `universal scaling' in terms of $\omega/\Delta_g$ (Fig.~\ref{fig:qpf_lma2}), which we find to be dominated by slow logarithmic tails just as for the metallic AIM, and for which the LMA provides analytic results (see e.g.\ Eq.~(\ref{eq:log_tails})). We naturally expect that these features will also show up in the dynamic and transport properties of the model at finite-$T$, a subject to which we will turn in a subsequent paper. 

Finally, the present work has been confined intentionally to the symmetric PAM relevant to the Kondo insulators. This is of course a special, albeit physically important limiting case of the asymmetric PAM with arbitrary conduction band filling $n_c$, which for $n_c \neq 1$ describes the metallic heavy fermion compounds; and extension of the LMA to encompass which is currently in hand. 
 
\begin{acknowledgement}
It is a pleasure to acknowledge many helpful discussions with R.\ Bulla, T.\ Pruschke and N.\ S.\ Vidhyadhiraja. We are also grateful for support from the Royal Society and the Indian National Science Academy (DEL, HRK), and the EPSRC and Leverhulme Trust (VES, DEL).
\end{acknowledgement}
   


\begin{thebibliography}{}
\bibitem{41}{This concurs with previous work on the symmetric PAM, see e.g.\ \cite{7,8,9,10,19,40}; and has been argued to be at variance with Nozi$\grave{\mathrm{e}}$res' `exhaustion physics' \cite{42}, characterised by two distinct low-energy scales, that has recently provoked considerable study \cite{8,11,12,17,43}.}
\end{thebibliography}
\end{document}